\documentclass[11pt]{article}
\usepackage{calc}
\usepackage{amsmath}
\usepackage{enumitem}
\usepackage[preprint]{latex/acl}
\usepackage{hyperref}
\usepackage{times}
\usepackage{latexsym}

\usepackage[T1]{fontenc}

\usepackage[utf8]{inputenc}

\usepackage{microtype}

\usepackage{graphicx}
\graphicspath{{../figures/}{./figures/}}
\usepackage{booktabs}

\title{\textsc{ParaCodex}: A Profiling-Guided Autonomous Coding Agent for Reliable Parallel Code Generation and Translation}

\author{
  \begin{tabular}{c}
    Erel Kaplan$^1$ \, Tomer Bitan$^1$ \, Lian Ghrayeb$^1$ \, Le Chen$^2$ \\
    Tom Yotam$^3$ \, Niranjan Hasabnis$^3$ \, Gal Oren$^{1,4}$
  \end{tabular} \\
  $^1$Technion -- Israel Institute of Technology, Haifa, Israel \\
  $^2$Argonne National Laboratory, Lemont, USA \\
  $^3$Code Metal, USA \\
  $^4$Stanford University, Stanford, USA\\
  \texttt{\{erel.kaplan, tomerbitan, lian.ghrayeb\}@campus.technion.ac.il} \\
  \texttt{lechen@anl.gov}, \texttt{\{tom, niranjan\}@codemetal.ai}, \\ \texttt{galoren@stanford.edu}
}

\usepackage{tikz}
\usetikzlibrary{calc}
\usepackage{xcolor}

\usepackage{pgfplots}
\pgfplotsset{compat=1.18}

\usepackage{capt-of}

\makeatletter
\AtBeginDocument{%
  \let\@oldmaketitle\@maketitle
  \renewcommand{\@maketitle}{%
    \@oldmaketitle
    \begin{center}
      \includegraphics[width=\linewidth]{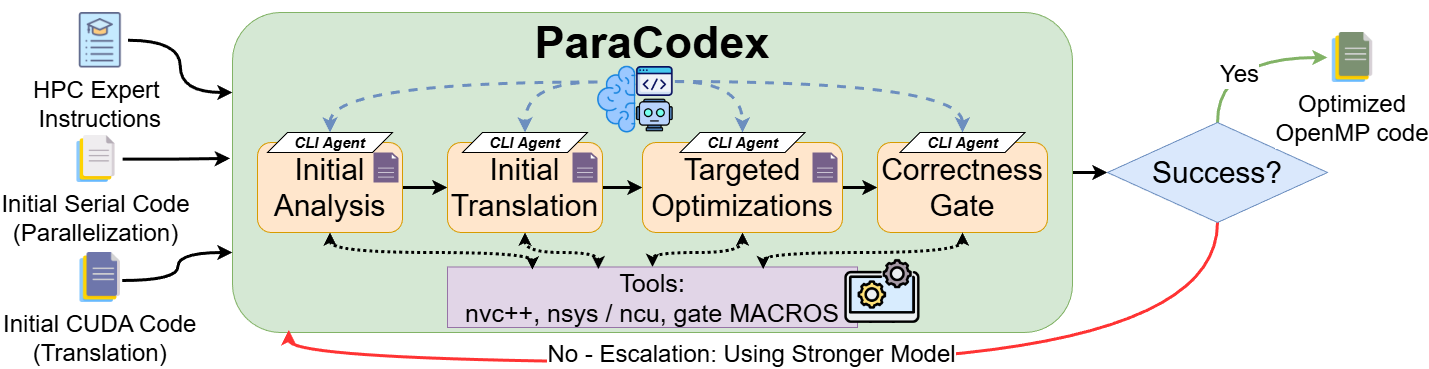}
      \captionof{figure}{\textbf{\textsc{ParaCodex}: An autonomous LLM agent that parallelizes and migrates code.} The core idea is to turn generation into an artifact-driven, tool-verified agentic workflow: the agent first extracts bottlenecks and a data-mapping plan, then proposes patch-level OpenMP \texttt{target} edits, and finally uses compiler/tests as a hard correctness gate and profiler traces as a performance signal. By separating \emph{what to move} (data planning) from \emph{what to offload} (kernel translation) and closing the loop with measurement, \textsc{ParaCodex} reduces brittle one-shot outputs and makes iterations reproducible via structured plans, logs, and profiles.}

      \label{fig:paracodex-flow}
    \end{center}
    \vspace{10pt} 
  }%
}
\makeatother

\begin{document}
\maketitle

\newcommand{\PC}{\textsc{ParaCodex}}

\begin{abstract}
Parallel programming is central to HPC and AI, but producing code that is correct and fast remains challenging, especially for OpenMP GPU offload, where data movement and tuning dominate. \emph{Autonomous coding agents} can compile, test, and profile on target hardware, but outputs are brittle without domain scaffolding.

We present \textsc{ParaCodex}, an HPC-engineer workflow that turns a Codex-based agent into an autonomous OpenMP GPU offload system using staged hotspot analysis, explicit data planning, correctness gating, and profiling-guided refinement. We evaluate translation from serial CPU kernels to OpenMP GPU offload kernels on HeCBench, Rodinia, and NAS\@. After excluding five kernels, \textsc{ParaCodex} succeeded on all 31 valid kernels. The generated kernels improved GPU time over reference OpenMP implementations in 25/31 cases, achieving geometric-mean speedups of 3$\times$ on HeCBench and 5$\times$ on Rodinia, and outperforming a zero-shot Codex baseline on all suites. We also evaluate CUDA$\rightarrow$OpenMP offload translation on ParEval, where \textsc{ParaCodex} maintains high compilation and validation rates in code-only and end-to-end settings~\footnote{\textsc{ParaCodex} repository: 
\url{https://github.com/Scientific-Computing-Lab/ParaCodex}}.
\end{abstract}

\section{Introduction}
Parallel programming is central to modern high-performance computing, but producing parallel implementations that are both correct and fast remains arduous. In practice, developers face two recurring needs: \emph{introducing} parallelism into serial kernels to exploit CPUs and GPUs, and \emph{migrating} existing parallel code between programming models to improve portability and maintainability.

Crucially, this work studies \emph{tool-using LLM code autonomous agents} for program transformation: systems that do not stop at text-only code generation, but iteratively compile, test, and profile their outputs on target hardware, using structured intermediate artifacts (e.g., hotspot analyses and data plans) and environment feedback as first-class signals for refinement~\citep{gottschlich2018three}.

Classical automation tools like polyhedral compilers~\citep{bondhugula2008pluto,doerfert2015pollyspolyhedralschedulingpresence} reduce manual effort but struggle with non-affine loops and ambiguous pointer aliasing, often requiring expert intervention to inject pragmas or resolve dependencies~\citep{harel2020source,mosseri2020compar}. Conversely, while Large Language Models (LLMs) can generate plausible OpenMP directives, they lack the \emph{system-level awareness} to manage data movement efficiently~\citep{nichols2024hpccoder,bitan2025unipar,dearing2025lassi}. A common failure mode is ``thrashing,'' where an LLM correctly parallelizes a loop but fails to map a helper function or hoist data allocations, causing the runtime to implicitly copy arrays back and forth between CPU and GPU at every iteration, yielding code that is functionally correct but much slower than serial execution.

Unlike text-only code generation, \emph{autonomous coding agents}~\citep{li2025rise} integrate code models with tool use in terminals or IDEs, enabling iterative editing, compilation, and testing directly on the target machine~\citep{yang2024swe}. This shifts the setting from offline code completion to \textit{in-situ}, feedback-driven development. We focus on OpenMP GPU offloading~\citep{deakin2023programming} because it addresses the ``ultimate goal'' of modern parallel computing with one of the most used parallel APIs~\citep{valero2025chathpc,kadosh2023quantifying}: exploiting the massive performance advantage of GPUs over CPUs, which has become ubiquitous over the last decade~\citep{top500_2025_11}. By targeting OpenMP offloading, we tackle a triple challenge: (i) maintaining portability across heterogeneous systems, (ii) preserving CPU fallback (since OpenMP offloads are optional), and (iii) raising the bar for agentic reasoning, requiring the agent to correctly manage device memory and kernel launches, not just thread-level parallelism. We evaluate two input settings: \emph{translation} (serial$\rightarrow$OpenMP) and \emph{migration} (CUDA$\rightarrow$OpenMP), among the most important tasks in the field.~\citep{chen2024landscape}.

These limitations motivate the design of \textsc{ParaCodex}, a Codex-based agentic workflow~\citep{horikawa2025agentic,vangala2025ai} that mirrors the iterative process of an HPC engineer~\citep{mondesire2025automating}. To prevent premature parallelization of unsafe loops (a classical tool weakness), \textsc{ParaCodex} enforces Stage 1 (Analysis). To avoid data thrashing, Stage 2 (Data Planning) explicitly structures memory residency before code generation. Finally, to catch performance regressions that static analysis misses, Stage 3 (Profiling) closes the loop with ground-truth measurement~\citep{peng2024perfcodegen}. Unlike approaches that stop once tests pass, \textsc{ParaCodex} explicitly optimizes for performance via feedback.

We evaluate \textsc{ParaCodex} on 36 serial$\rightarrow$OpenMP tasks drawn from HeCBench~\citep{jin2023hecbench} (23), Rodinia~\citep{che2009rodinia} (7), and NAS~\citep{bailey1991nas, fridman2025openacc} (6). Together, these widely used suites span micro-kernels through end-to-end scientific benchmarks, with provided implementations ranging from optimized to highly tuned. Using GPU time measured by NVIDIA Nsight Systems (kernel execution plus transfers), \textsc{ParaCodex} achieves valid GPU offload on 31/36 kernels (86\%; 5 kernels excluded, detailed in \S\ref{sec:accounting}), produces correct implementations for all 31 valid kernels, and improves GPU time over the provided OpenMP implementations in 25/31 cases (80\%), with geometric-mean GPU-time speedups of 3$\times$ on HeCBench, 5.1$\times$ on Rodinia, and 1.08$\times$ on the highly tuned NAS suite. As a separate generalization study, we evaluate CUDA$\rightarrow$OpenMP migration on four ParEval~\citep{nichols2024large} kernels, where \textsc{ParaCodex} maintains high compilation and validation rates under both a fixed build environment (\emph{code-only}) and an end-to-end setting where the agent must also construct the build flow (\emph{overall}) (\S\ref{sec:experiments}).

\paragraph{Key Question.} \textit{To what extent can a tool-using autonomous coding agent translate and migrate parallel code under automated build-and-test validation, and use profiling feedback to narrow the performance gap to reference implementations?}

\paragraph{Research Questions.} To explore this space we pose four guiding questions:
\begin{enumerate}[nosep,topsep=0pt,leftmargin=*,labelsep=.2em,widest*=9]
    \item \textbf{RQ1 -- Baseline feasibility:} How effectively can contemporary agentic LLMs translate serial code into correct OpenMP GPU offload without profiling-guided iteration?
    \item \textbf{RQ2 -- Benefit of specialization:} Does a staged, expert-seeded workflow with structured intermediate artifacts improve robustness and performance?
    \item \textbf{RQ3 -- Performance attainment:} To what extent can profiling-in-the-loop refinement approach or exceed the provided OpenMP implementations across diverse benchmarks, and where do regressions persist?
    \item \textbf{RQ4 -- Task generalization:} Does the same workflow generalize from serial$\rightarrow$OpenMP translation to CUDA$\rightarrow$OpenMP migration?
\end{enumerate}

\paragraph{Contributions.} We make three contributions:
\begin{enumerate}[nosep,topsep=0pt,leftmargin=*,labelsep=.2em,widest*=9]
    \item We systematize HPC engineering practice into a reusable agentic pattern: from hotspot analysis via a loop taxonomy through data residency and transfer strategy selection to profile-guided tuning, connected via Makefile-based correctness gates (RQ2, RQ3).
    \item We design a comprehensive evaluation protocol for serial$\rightarrow$OpenMP agentic parallelization across HeCBench, Rodinia, and NAS under a unified build-and-test harness and a zero-shot Codex baseline, quantifying both robustness and GPU-time performance relative to expert implementations (RQ1, RQ3).
    \item We demonstrate generalization to CUDA$\rightarrow$OpenMP translation on ParEval, and introduce bypass-detection analysis that surfaces pseudo-offload cases and suggests harness design strategies to enforce device-side work (RQ4).
\end{enumerate}

\section{Related Work}
We organize prior work from static and learned parallelization methods, through LLM-based parallel code generation and translation, to autonomous coding agents and feedback-driven self-refinement. We distinguish \textsc{ParaCodex} from prior work along three axes: (1) Compile/test repair (UniPar, LASSI) fixes bugs but ignores speed; (2) Performance feedback (PerfCodeGen, STOP) optimizes runtime but lacks rigorous correctness gates; and (3) Artifact-driven planning -- absent in prior agents -- externalizes reasoning before coding. \textsc{ParaCodex} is, to our knowledge, the first system to integrate all three axes.

\paragraph{Compilers and Learned Parallelization.} Compiler research explored automatic loop transformations via dependence analysis and the polyhedral model~\citep{bondhugula2008pluto,doerfert2015pollyspolyhedralschedulingpresence}. Such tools excel at affine loop nests but typically require expert intervention. Survey work analyzed source-to-source OpenMP compilers, highlighting pitfalls and variability~\citep{harel2020source,mosseri2020compar}.
Recent efforts apply learned cost models, RL, and autotuning to optimization~\citep{chen2019learningoptimizetensorprograms,ahn2019reinforcementlearningadaptivesampling,wu2023autotuningapachetvmbasedscientific,ding2023hidettaskmappingprogrammingparadigm,merouani2025looperlearnedautomaticcode}, relying on iterative search and measurement.
ML techniques predict parallelization opportunities and suggest OpenMP directives~\citep{maramzin2019parallelloop,chen2023learningopenmp,harel2023learning,kadosh2024ompar,harel2025pragformer}, but stop short of generating complete code.

\paragraph{LLMs for Parallel Code.}
Domain-focused LLMs (HPC-Coder, OMPGPT, MonoCoder) specialize in parallel kernels~\citep{nichols2024hpccoder,chen2024ompgpt,kadosh2024monocoder,schneider2024mpirigen,kadosh2023scope,chen2024landscape}. Broader models (CodeGeeX, Meta's LLM Compiler) extend coverage~\citep{zheng2024codegeexpretrainedmodelcode,cummins2024metalargelanguagemodel}. Transformer advisors (OMPify, MPI-rical) provide data-driven guidance~\citep{kadosh2023advising,schneider2023mpi,schneider2024mpirigen}. However, these typically deliver single-shot outputs without iterative feedback.

\paragraph{Autonomous Coding Agents.}
Recent work describes \emph{autonomous coding agents}: systems pairing code models with developer tools in iterative loops~\citep{li2025rise}. Examples include CLI assistants and IDE integrations (GitHub Copilot, OpenAI Codex, Cursor). This is relevant for parallel code, where correctness requires build-and-test harnesses and performance depends on concrete hardware. Codex has been evaluated on code-generation benchmarks with strong performance.\footnote{\url{https://openai.com/index/introducing-gpt-5-2-codex/}} We build \textsc{ParaCodex} on Codex. Tool access alone is insufficient: without domain structure, workflows can be brittle. Prior pipelines emphasize compilation repair and correctness~\citep{bitan2025unipar,dearing2025lassi, chen2025beyond}, whereas \textsc{ParaCodex} operationalizes an HPC-engineer with explicit artifacts, correctness gates, and profiling-guided optimization.

\paragraph{LLM-Based Translation Pipelines.}
Frameworks such as UniPar and LASSI study LLMs for translating parallel code, including serial$\rightarrow$OpenMP~\citep{bitan2025unipar,dearing2025lassi,chen2025beyond}. They implement \emph{compile/test-driven repair}: compiler errors and test failures trigger iterative fixes, improving compilation and correctness. However, they lack (i) \emph{execution/performance feedback} via profiling to guide optimization, and (ii) \emph{artifact-driven structured plans} that externalize reasoning before code modification. Complementary work quantifies LLM capability on HPC kernel generation~\citep{godoy2023evaluationopenaicodexhpc,valerolara2023comparingllama2gpt3llms,cui2025largelanguagemodelsunderstand,bolet2025can}.

\paragraph{Self-Refinement and Feedback-Driven Optimization.}
Agentic research explores deliberate search and refinement for program synthesis~\citep{yao2023treethoughtsdeliberateproblem,shinn2023reflexion,singhal2025llmermsampleefficientprogramlearning,madaan2023selfrefine,jimenez2024swebenchlanguagemodelsresolve}.
PerfCodeGen demonstrates execution-driven refinement by feeding performance metrics back~\citep{peng2024perfcodegen}, and Self-Taught Optimizer (STOP) scaffolds LLMs to recursively optimize programs~\citep{zelikman2023stop}. These systems show the value of \emph{execution feedback}, but do not enforce \emph{correctness gating} at each stage or emit \emph{structured intermediate artifacts} (e.g., data plans, optimization plans) that anchor reasoning to domain constraints. \textsc{ParaCodex} combines all three mechanisms -- compile/test repair, profiling feedback, and artifact-driven planning -- within a unified HPC-engineering workflow.

\section{\textsc{ParaCodex} Agent Overview}

\textsc{ParaCodex} is a guided agentic workflow in which the model iteratively uses standard development tools. The agent first analyzes the input program to identify performance-critical loops or kernels, then generates an OpenMP GPU version. It compiles and executes the code against the serial reference, using compilation errors and numerical mismatches as structured feedback to drive targeted fixes until correctness is achieved.

The workflow then shifts to performance tuning. The agent profiles execution with a profiler and applies focused optimizations to further improve performance. In combination, correctness-driven repair and profile-guided tuning enable robust automated optimization.

\subsection{Baseline Comparison System.}
\label{sec:baseline-definition}
Throughout this paper, we use the term \emph{baseline} to refer to a zero-shot Codex CLI agent~\footnote{\url{https://platform.openai.com/docs/models/gpt-5.1-codex}} that directly translates the input program into OpenMP GPU offload in a single pass. Both baseline and \textsc{ParaCodex} are evaluated under the same build and correctness harness; \textsc{ParaCodex} additionally uses staged prompts and profiling-guided refinement. This setup keeps the underlying model fixed and isolates the impact of workflow scaffolding. The baseline is explained in more detail in App.~\ref{app:baseline-prompt}.

\subsection{Pipeline Stages}
\textsc{ParaCodex} executes a three-stage workflow as depicted in \autoref{fig:paracodex-flow}. (See App.~\ref{app:stage-restructuring} for details on suite-specific adaptations.)

\begin{enumerate}[nosep,topsep=0pt,leftmargin=*,labelsep=.2em,widest*=9]
    \item \textbf{Analysis: Loop classification and data characterization.} Inspects input code to identify and rank candidate loops by computational weight, defined as the estimated operation count (iterations $\times$ ops/iteration) to distinguish $O(N)$ critical paths from setup code. It generates \texttt{analysis.md}, which classifies loops using a taxonomy (Types A--G) and records data properties, dependencies, and hazards (App.~\ref{app:analysis-data-opt}). \emph{Rationale:} Externalizing reasoning before code modification reduces chances of premature transformation of loops with hidden hazards (e.g., reductions, recurrences) that would silently break correctness.
    
    \item \textbf{GPU offloading + data strategy: Correctness with explicit data plan.} 
    Before inserting pragmas, the agent selects a data-management strategy based on the loop taxonomy: (A) \emph{Scoped Regions} use explicit \texttt{target data} directives for standard kernels to minimize transfers; (B) \emph{Asynchronous Pipelines} use \texttt{nowait} for overlapping computation and communication; (C) \emph{Global Device State} allocates persistent device memory (\texttt{omp\_target\_alloc}) for iterative solvers to avoid repeated mapping. 
    It writes \texttt{data\_plan.md} specifying array residency and function offload status, prioritizing \emph{CRITICAL} loops (those with high computational weight) for offload.
    This explicit planning step reduces the chances of common ``thrashing'' pitfalls, where unmapped helper functions trigger implicit host-device transfers at every iteration.
    
    \item \textbf{Performance tuning: Profile-guided optimization.}
    After functional validation, the agent profiles the program with NVIDIA Nsight Systems, extracting kernel times, transfer volumes, and API overheads. It generates an \texttt{optimization\_plan.md} that ranks bottlenecks and prescribes targeted fixes (e.g., fusing adjacent kernels to reduce launch latency, collapsing loops to increase occupancy, hoisting transfers out of iterative loops). A strict revert-on-regression policy is enforced: if GPU time worsens by more than 10\%, the change is automatically rolled back.

    \item \textbf{Correctness gating.} Between each stage, a \emph{correctness-gate agent} is invoked. The agent instruments the code with \texttt{gate.h} -- a lightweight header that computes checksums and norms at key program points -- to localize the divergence and drive the repairs. It then compiles and runs the candidate code against the input serial code using a manually-written Makefile-based harness. Then, it corrects the code if the validation fails (via exit code or numerical mismatch). (App.~\ref{app:analysis-data-opt:correctness}). 

\end{enumerate}

For space purposes, a concrete example showing the end-to-end workflow of hotspot analysis, data plan, and optimization is kept in App.~\ref{app:running-example}.

\subsection{Tooling Integration}
The pipeline uses external tools accessible to the agent. Compiler diagnostics from NVIDIA HPC SDK surface issues; Makefile-based harnesses validate correctness; NVIDIA Nsight Systems provides performance diagnosis; system metadata is read from \texttt{system\_info.txt}. The design combines LLM-driven generation with deterministic tool feedback for iterative repair and tuning.

\subsection{Prompt Engineering Strategy}
\textsc{ParaCodex} uses multi-stage prompting that structures the model's reasoning across analysis, translation, and optimization. Each stage is driven by a specific template that constrains the output space while retaining domain flexibility.

\begin{enumerate}[nosep,topsep=0pt,leftmargin=*,labelsep=.2em,widest*=9]
\item\textbf{Analysis Phase.}
The model produces a structured hotspot analysis (App.~\ref{app:analysis-data-opt:analysis}) that characterizes loops using a taxonomy (Types A--G) capturing parallelization constraints. This prompt converts qualitative reasoning into a discrete intermediate representation, revealing any hazards before transformation.

\item\textbf{Translation Phase.}
Using the structured analysis, the prompt maps each loop class to a restricted set of valid OpenMP constructs (App.~\ref{app:analysis-data-opt:data}), minimizing mapping errors like omitted reductions.

\item\textbf{Optimization Phase.}
This stage implements measurement-guided search (App.~\ref{app:analysis-data-opt:opt}). The prompt pairs Nsight Systems profiles with optimization levers to apply targeted modifications, retaining only those that improve performance.

\end{enumerate}

\section{Experimental Evaluation}
\label{sec:experiments}

\subsection{Experimental Setup}
\paragraph{Environment and Protocol.} We use an NVIDIA RTX 4060 Laptop GPU (8GB) and i9-13905H CPU with NVIDIA HPC SDK 25.7 (\texttt{nvc++}). We report \emph{GPU time} measured via NVIDIA Nsight Systems (v2024.5), computed by summing CUDA kernel time (\texttt{cuda\_gpu\_kern\_sum}) and device transfer time (\texttt{cuda\_gpu\_mem\_time\_sum}), excluding CUDA API and runtime overhead. We perform three profiling runs per configuration and report the mean. 

\paragraph{GPU Time Metric.}
\label{gpu-time}
We define \emph{GPU time} as the sum of CUDA kernel execution time and device memory transfer time from NVIDIA Nsight Systems, excluding CUDA API overhead. This captures computational work and data movement on the device. The implications are that (i) GPU time represents device-side performance when offload is substantial; (ii) it can diverge from wall-clock time when API/host overhead dominates; (iii) CPU-fallback cases yield misleadingly low GPU times as detailed in \S\ref{sec:accounting}; (iv) if there are no kernel launches at all, the profiler automatically detects it.

\definecolor{BlueLow}{RGB}{247,251,255}   
\definecolor{BlueHigh}{RGB}{8,48,107}     
\definecolor{GreenLow}{RGB}{247,252,245}  
\definecolor{GreenHigh}{RGB}{0,68,27}     

\newcommand{\CW}{0.46cm}   
\newcommand{\CH}{0.42cm}   
\newcommand{\HGap}{0.25cm} 
\newcommand{\VGap}{0.40cm} 
\newcommand{\TitleY}{0.30cm} 

\newcommand{\TitleFont}{\scriptsize\bfseries}
\newcommand{\CellFont}{\scriptsize}
\newcommand{\AxisFont}{\scriptsize}
\newcommand{\TickFont}{\scriptsize}
\newcommand{\CbarFont}{\scriptsize}

\newcommand{\SetCellColor}[2]{
  \pgfmathsetmacro{\pp}{100*(#2)}
  \colorlet{cellfill}{#1High!\pp!#1Low}
}

\newcommand{\DrawHeatmap}[7]{%
  \begin{scope}[shift={(#2,#3)}]
    \ifx#4\empty\else
      \node[font=\TitleFont] at ({#7}, \TitleY) {#4};
    \fi

    \foreach \r/\c/\v/\lab in {#6} {
      \SetCellColor{#1}{\v}
      \coordinate (P) at (\c*\CW, -\r*\CH);
      \fill[cellfill] (P) rectangle ++(\CW,-\CH);
      \draw[white, line width=0.4pt] (P) rectangle ++(\CW,-\CH);
      \pgfmathparse{\v > 0.5 ? 1 : 0}
      \ifnum\pgfmathresult=1
        \node[font=\CellFont, text=white] at ($(P)+(.5*\CW,-.5*\CH)$) {\lab};
      \else
        \node[font=\CellFont, text=black] at ($(P)+(.5*\CW,-.5*\CH)$) {\lab};
      \fi
    }

    \ifnum#5=1
      \foreach \r/\name in {0/nanoXOR,1/microXORH,2/microXOR,3/XSBench}{
        \node[anchor=east,font=\AxisFont] at (0, -\r*\CH-0.5*\CH) {\name};
      }
    \fi
  \end{scope}
}

\newcommand{\DrawXLabelsFive}[2]{
  \begin{scope}[shift={(#1,#2)}]
    \foreach \c/\t in {
      0/gemini-1.5,
      1/gpt-4o-mini,
      2/o4-mini,
      3/llama-3.3,
      4/qwen2-32b
    }{
      \node[font=\TickFont, rotate=45, anchor=north east] at (\c*\CW+0.05cm, 0) {\t};
    }
  \end{scope}
}

\newcommand{\DrawXLabelOne}[2]{
  \begin{scope}[shift={(#1,#2)}]
    \node[font=\TickFont, rotate=45, anchor=north east] at (0.05cm, 0) {gpt-5.1-codex-mini};
  \end{scope}
}

\newcommand{\DrawXLabelBaseline}[2]{
  \begin{scope}[shift={(#1,#2)}]
    \node[font=\TickFont, rotate=45, anchor=north east] at (0.05cm, 0) {gpt-5.1-codex-mini};
  \end{scope}
}

\newcommand{\DrawColorbar}[5]{%
  \begin{scope}[shift={(#2,#3)}]
    \pgfmathsetlengthmacro{\CbarMid}{0.5*#4}
    \shade[top color=#1High, bottom color=#1Low] (0,0) rectangle (0.25cm,#4);
    \draw (0,0) rectangle (0.25cm,#4);
    \foreach \t/\lab in {0/0,0.5/0.5,1/1}{
      \pgfmathsetlengthmacro{\CbarY}{\t*#4}
      \draw (0.25cm, \CbarY) -- (0.30cm, \CbarY);
      \node[font=\AxisFont, anchor=west] at (0.32cm, \CbarY) {\lab};
    }
    \ifx#5\empty\else
      \node[font=\AxisFont, rotate=90] at (0.85cm, \CbarMid) {#5};
    \fi
  \end{scope}
}

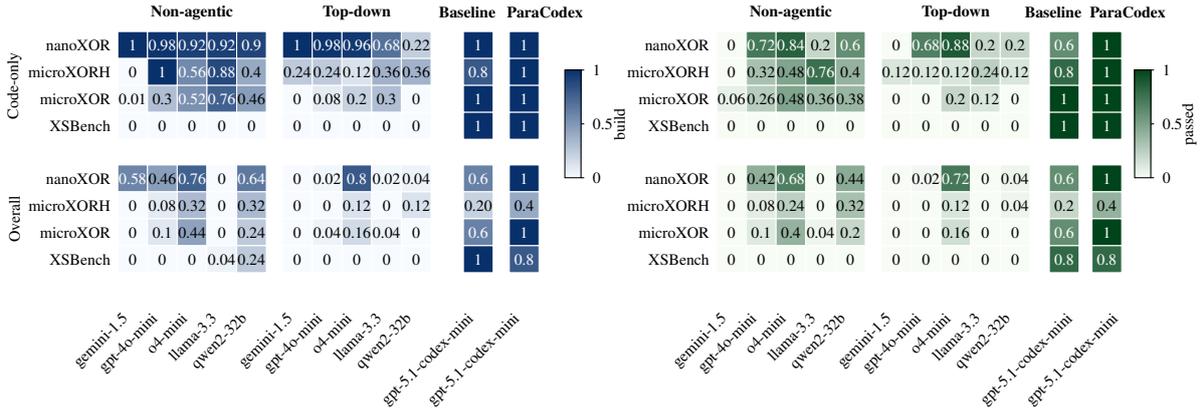
\begin{figure*}[!htbp]
  \centering
  \small
\begin{tikzpicture}[scale=0.85, every node/.style={transform shape}]

\pgfmathsetlengthmacro{\Wfive}{5*\CW}
\pgfmathsetlengthmacro{\Wone}{1*\CW}
\pgfmathsetlengthmacro{\Hfour}{4*\CH}
\pgfmathsetlengthmacro{\HalfHfour}{0.5*\Hfour}
\pgfmathsetlengthmacro{\YCodeOnly}{-\HalfHfour}
\pgfmathsetlengthmacro{\YOverall}{-(\Hfour+\VGap+\HalfHfour)}

\pgfmathsetlengthmacro{\XLabelY}{-(2*\Hfour+\VGap+0.45cm)}

\node[rotate=90, font=\AxisFont] at (-1.6cm, \YCodeOnly) {Code-only};
\node[rotate=90, font=\AxisFont] at (-1.6cm, \YOverall) {Overall};

\DrawHeatmap{Blue}{0}{0}{Non-agentic}{1}{
  0/0/1/1,    0/1/0.98/0.98, 0/2/0.92/0.92, 0/3/0.92/0.92, 0/4/0.90/0.9,
  1/0/0/0,    1/1/1/1,       1/2/0.56/0.56, 1/3/0.88/0.88, 1/4/0.40/0.4,
  2/0/0.01/0.01, 2/1/0.30/0.3, 2/2/0.52/0.52, 2/3/0.76/0.76, 2/4/0.46/0.46,
  3/0/0/0,    3/1/0/0,       3/2/0/0,       3/3/0/0,       3/4/0/0
}{2.5*\CW}
\DrawHeatmap{Blue}{\Wfive+\HGap}{0}{Top-down}{0}{
  0/0/1/1,    0/1/0.98/0.98, 0/2/0.96/0.96, 0/3/0.68/0.68, 0/4/0.22/0.22,
  1/0/0.24/0.24, 1/1/0.24/0.24, 1/2/0.12/0.12, 1/3/0.36/0.36, 1/4/0.36/0.36,
  2/0/0/0,    2/1/0.08/0.08, 2/2/0.20/0.2,  2/3/0.30/0.3,  2/4/0/0,
  3/0/0/0,    3/1/0/0,       3/2/0/0,       3/3/0/0,       3/4/0/0
}{2.5*\CW}
\DrawHeatmap{Blue}{2*\Wfive+3*\HGap}{0}{Baseline}{0}{
  0/0/1/1,
  1/0/0.8/0.8,
  2/0/1/1,
  3/0/1/1
}{0.13*\CW}
\DrawHeatmap{Blue}{2*\Wfive+3*\HGap+\Wone+\HGap}{0}{ParaCodex}{0}{
  0/0/1/1,
  1/0/1/1,
  2/0/1/1,
  3/0/1/1
}{1.2*\CW}

\DrawColorbar{Blue}{2*\Wfive+3*\HGap+2*\Wone+2*\HGap+0.15cm}{-(\Hfour + 1.5*\VGap)}{\Hfour}{build}

\DrawHeatmap{Blue}{0}{-(\Hfour+\VGap)}{}{1}{
  0/0/0.58/0.58, 0/1/0.46/0.46, 0/2/0.76/0.76, 0/3/0/0,    0/4/0.64/0.64,
  1/0/0/0,       1/1/0.08/0.08, 1/2/0.32/0.32, 1/3/0/0,    1/4/0.32/0.32,
  2/0/0/0,       2/1/0.10/0.1,  2/2/0.44/0.44, 2/3/0/0,    2/4/0.24/0.24,
  3/0/0/0,       3/1/0/0,       3/2/0/0,       3/3/0.04/0.04, 3/4/0.24/0.24
}{2.5*\CW}
\DrawHeatmap{Blue}{\Wfive+\HGap}{-(\Hfour+\VGap)}{}{0}{
  0/0/0/0,       0/1/0.02/0.02, 0/2/0.80/0.8,  0/3/0.02/0.02, 0/4/0.04/0.04,
  1/0/0/0,       1/1/0/0,       1/2/0.12/0.12, 1/3/0/0,       1/4/0.12/0.12,
  2/0/0/0,       2/1/0.04/0.04, 2/2/0.16/0.16, 2/3/0.04/0.04, 2/4/0/0,
  3/0/0/0,       3/1/0/0,       3/2/0/0,       3/3/0/0,       3/4/0/0
}{2.5*\CW}
\DrawHeatmap{Blue}{2*\Wfive+3*\HGap}{-(\Hfour+\VGap)}{}{0}{
  0/0/0.60/0.6,
  1/0/0.20/0.20,
  2/0/0.60/0.6,
  3/0/1/1
}{2.5*\CW}
\DrawHeatmap{Blue}{2*\Wfive+3*\HGap+\Wone+\HGap}{-(\Hfour+\VGap)}{}{0}{
  0/0/1/1,
  1/0/0.40/0.4,
  2/0/1/1,
  3/0/0.80/0.8
}{2.5*\CW}

\DrawXLabelsFive{0}{\XLabelY}
\DrawXLabelsFive{\Wfive+\HGap}{\XLabelY}
\DrawXLabelBaseline{2*\Wfive+3*\HGap}{\XLabelY}
\DrawXLabelOne{2*\Wfive+3*\HGap+\Wone+\HGap}{\XLabelY}

\pgfmathsetlengthmacro{\GX}{2*\Wfive+3*\HGap+2*\Wone+2*\HGap+2.5cm}

\DrawHeatmap{Green}{\GX}{0}{Non-agentic}{1}{
  0/0/0/0,    0/1/0.72/0.72, 0/2/0.84/0.84, 0/3/0.20/0.2,  0/4/0.60/0.6,
  1/0/0/0,    1/1/0.32/0.32, 1/2/0.48/0.48, 1/3/0.76/0.76, 1/4/0.40/0.4,
  2/0/0.06/0.06, 2/1/0.26/0.26, 2/2/0.48/0.48, 2/3/0.36/0.36, 2/4/0.38/0.38,
  3/0/0/0,    3/1/0/0,       3/2/0/0,       3/3/0/0,       3/4/0/0
}{2.5*\CW}
\DrawHeatmap{Green}{\GX+\Wfive+\HGap}{0}{Top-down}{0}{
  0/0/0/0,    0/1/0.68/0.68, 0/2/0.88/0.88, 0/3/0.20/0.2,  0/4/0.20/0.2,
  1/0/0.12/0.12, 1/1/0.12/0.12, 1/2/0.12/0.12, 1/3/0.24/0.24, 1/4/0.12/0.12,
  2/0/0/0,    2/1/0/0,       2/2/0.20/0.2,  2/3/0.12/0.12, 2/4/0/0,
  3/0/0/0,    3/1/0/0,       3/2/0/0,       3/3/0/0,       3/4/0/0
}{2.5*\CW}
\DrawHeatmap{Green}{\GX+2*\Wfive+3*\HGap-0.2cm}{0}{Baseline}{0}{
  0/0/0.60/0.6,
  1/0/0.8/0.8,
  2/0/1/1,
  3/0/1/1
}{0.13*\CW}
\DrawHeatmap{Green}{\GX+2*\Wfive+3*\HGap+\Wone+\HGap-0.25cm}{0}{ParaCodex}{0}{
  0/0/1/1,
  1/0/1/1,
  2/0/1/1,
  3/0/1/1
}{1.2*\CW}

\DrawColorbar{Green}{\GX+2*\Wfive+3*\HGap+2*\Wone+2*\HGap-0.3cm}{-(\Hfour + 1.5*\VGap)}{\Hfour}{passed}

\DrawHeatmap{Green}{\GX}{-(\Hfour+\VGap)}{}{1}{
  0/0/0/0,    0/1/0.42/0.42, 0/2/0.68/0.68, 0/3/0/0,    0/4/0.44/0.44,
  1/0/0/0,    1/1/0.08/0.08, 1/2/0.24/0.24, 1/3/0/0,    1/4/0.32/0.32,
  2/0/0/0,    2/1/0.10/0.1,  2/2/0.40/0.4,  2/3/0.04/0.04, 2/4/0.20/0.2,
  3/0/0/0,    3/1/0/0,       3/2/0/0,       3/3/0/0,    3/4/0/0
}{2.5*\CW}
\DrawHeatmap{Green}{\GX+\Wfive+\HGap}{-(\Hfour+\VGap)}{}{0}{
  0/0/0/0,    0/1/0.02/0.02, 0/2/0.72/0.72, 0/3/0/0,    0/4/0.04/0.04,
  1/0/0/0,    1/1/0/0,       1/2/0.12/0.12, 1/3/0/0,    1/4/0.04/0.04,
  2/0/0/0,    2/1/0/0,       2/2/0.16/0.16, 2/3/0/0,    2/4/0/0,
  3/0/0/0,    3/1/0/0,       3/2/0/0,       3/3/0/0,    3/4/0/0
}{2.5*\CW}
\DrawHeatmap{Green}{\GX+2*\Wfive+3*\HGap-0.2cm}{-(\Hfour+\VGap)}{}{0}{
  0/0/0.60/0.6,
  1/0/0.2/0.2,
  2/0/0.60/0.6,
  3/0/0.80/0.8
}{0.13*\CW}
\DrawHeatmap{Green}{\GX+2*\Wfive+3*\HGap+\Wone+\HGap-0.25cm}{-(\Hfour+\VGap)}{}{0}{
  0/0/1/1,
  1/0/0.40/0.4,
  2/0/1/1,
  3/0/0.80/0.8
}{1.2*\CW}

\DrawXLabelsFive{\GX}{\XLabelY}
\DrawXLabelsFive{\GX+\Wfive+\HGap}{\XLabelY}
\DrawXLabelBaseline{\GX+2*\Wfive+3*\HGap}{\XLabelY}
\DrawXLabelOne{\GX+2*\Wfive+3*\HGap+\Wone+\HGap}{\XLabelY}

\end{tikzpicture}
\vspace{-0.3cm}
\caption{\textbf{CUDA$\rightarrow$OpenMP translation quality on ParEval across scoring regimes.}
Compilation (blue) and validation (green) success rates are shown for both the baseline Codex model and \textsc{ParaCodex} under the ParEval protocol. Kernels are scored under (i) a fixed build environment (\emph{code-only}) and (ii) a setting where the system must also construct the build flow (\emph{overall}). The heatmap layout mirrors the original ParEval presentation. Both models perform strongly; \textsc{ParaCodex} matches or slightly exceeds the baseline, indicating that the agentic workflow generalizes to cross-API translation.\vspace{-0.5cm}}

  \label{fig:pareval}
\end{figure*}

\paragraph{Model Selection and Escalation.} We use \texttt{gpt-5.1-codex-mini} as the primary model for all kernels in both \textsc{ParaCodex} and the baseline. For the NAS FT kernel, which failed with the base model, we escalated to \texttt{gpt-5.1-codex-max} for both \textsc{ParaCodex} and the baseline. In all experiments, the agent operated without internet access and did not use web search at any stage.

\begin{table}[!t]
\centering
\small
\begin{tabular}{lccc}
\toprule
\textbf{Suite} & \textbf{Attempted} & \textbf{Valid} & \textbf{Improved} \\
\midrule
HeCBench & 23 & 21 & 18/21 (86\%) \\
Rodinia & 7  & 6  & 4/6 (67\%) \\
NAS     & 6  & 4  & 3/4 (75\%) \\
ParEval & 4  & 4  & -- \\
\midrule
\textbf{serial$\rightarrow$OpenMP} & \textbf{36} & \textbf{31} & \textbf{25/31 (80\%)} \\
\bottomrule
\end{tabular}
\caption{\textbf{Result accounting across suites.}
\emph{Attempted:} kernels tried; \emph{Valid GPU:} compile + pass correctness with substantial device execution; \emph{Improved:} \textsc{ParaCodex} reduces GPU time among valid GPU kernels. ParEval reports CUDA$\rightarrow$OpenMP separately.}
\label{tab:accounting}
\end{table}

\paragraph{Benchmarks and Baselines.} We use ParEval~\citep{nichols2024large}, HeCBench~\citep{jin2023hecbench}, Rodinia~\citep{che2009rodinia}, and NAS~\citep{bailey1991nas, fridman2025openacc}. From HeCBench, our evaluation considers 23 kernels that have been previously benchmarked in prior work. These consist of the 10 kernels analyzed in LASSI~\citep{dearing2025lassi} together with 13 kernels from the ParaTrans dataset utilized in UniPar~\citep{bitan2025unipar}. \emph{Reference implementations:} For all suites, the reference code is the provided OpenMP GPU-offload implementation from each benchmark; we compare \textsc{ParaCodex}'s generated OpenMP against those OpenMP baselines. \emph{Baseline for} \textsc{ParaCodex}: We compare \textsc{ParaCodex} against a zero-shot Codex baseline that receives the input code and a single prompt, as further detailed in \S\ref{sec:baseline-definition}. Both systems use the same constraints: identical build harness, correctness checks, profiling tools, and GPU/compiler environment. The key difference is the \emph{workflow}.
For each benchmark, we report speedup relative to reference as $T_{\text{ref}} / T_{\text{pc}}$, where values greater than one indicate lower execution time.

\subsection{Success Rate and Bypass Detection}
\label{sec:accounting}

\autoref{tab:accounting} provides a comprehensive accounting of kernel counts, exclusions, and success rates. We evaluate 36 serial$\rightarrow$OpenMP translation tasks (HeCBench: 23, Rodinia: 7, NAS: 6). We exclude 5 kernels from the final performance set: two from NAS (\textit{BT}, \textit{LU}) due to multi-file complexity; one from Rodinia (\textit{srad}) due to CPU-only reference implementation; and two from HeCBench (\textit{pathfinder}, \textit{particlefilter}) due to GPU-offload bypass (see App.~\ref{app:bypass}). This leaves 31 valid GPU-offload kernels.


Bypass was observed in 2/23 HeCBench kernels for \textsc{ParaCodex}, where implementations compile and pass correctness checks but execute the primary computation on the host CPU\@. We exclude bypass kernels from GPU-time statistics and provide detailed analysis in App.~\ref{app:bypass}.

\paragraph{Token Budget and Reproducibility.}
\textsc{ParaCodex} averages 837{,}701 tokens per kernel (1.11$\times$ baseline). This cost reflects a deliberate prioritization of \emph{reproducibility over efficiency}. Rather than brute-force sampling, we operate the agent in a non-interactive automation mode where full build logs and profiler traces are fed back to verify every step deterministically. This ensures the entire engineering loop is auditable and replayable without human intervention. To further support reproducibility, almost all experiments were run using \texttt{gpt-codex-5.1-mini} with the lowest reasoning effort, ensuring that independent researchers can replicate the outputs under the same configuration. Notably, the zero-shot baseline consumes comparable tokens (755{,}417) under the same harness, confirming that the cost stems from the robust tool-verified environment -- which accumulates context -- rather than the multi-stage logic itself (App.~\ref{app:token-count}).






\subsection{Performance Analysis of {\textsc{ParaCodex}}}
\definecolor{aclrose}{RGB}{204, 102, 119}
\definecolor{aclgold}{RGB}{221, 204, 119}

\begin{figure*}[!htbp]
\centering
\begin{tikzpicture}

\begin{axis}[
    at={(0,0)},
    anchor=south west,
    ybar=0pt,
    ymode=log,
    log basis y=10,
    width=0.95\textwidth,
    height=3.2cm,          
    bar width=4pt,
    enlarge x limits=0.03,
    symbolic x coords={
        ace-omp, atomicCost-omp, bsearch-omp, colorwheel-omp, convolution1D-omp, 
        dense-embedding-omp, entropy-omp, geodesic-omp, heat-omp, jacobi-omp, 
        keogh-omp, layout-omp, matrix-rotate-omp, michalewicz-omp, mixbench-omp, 
        pool-omp, randomAccess-omp, 
        stddev-omp, stencil1d-omp, vanGenuchten-omp, winograd-omp
    },
    xtick=data,
    xticklabels={
        ace, atomic-cost, bsearch, colorwheel, conv-1D, 
        dense-embedding., entropy, geodesic, heat, jacobi, 
        keogh, layout, matrix-rotate, michalewicz, mixbench, 
        pool, random-access, 
        std-dev, stencil-1D, van-genuchten, winograd
    },
    x tick label style={
        font=\tiny,         
        rotate=45,
        anchor=east
    },
    ymin=1, ymax=100000000, 
    ytick={1, 100, 10000, 1000000,100000000},
    yticklabels={1, 100, 10k, 1M, 100M},
    ylabel={Time ($\mu$s log$_{10}$)}, 
    ylabel style={font=\footnotesize},
    grid=both,
    major grid style={dashed,gray!40},
    minor grid style={dotted,gray!20},
    axis x line*=bottom,
    axis y line*=left,
    title={\textbf{HeCBench}},
    title style={font=\footnotesize, align=center},
    legend style={
        at={(0.16, 1.12)}, 
        anchor=south,
        legend columns=3,
        font=\footnotesize,
        draw=none
    }
]

\addplot[fill=black!30, draw=black!80, bar shift=-4pt] coordinates {
    (ace-omp, 1568730) (atomicCost-omp, 25308330) (bsearch-omp, 18) 
    (colorwheel-omp, 5) (convolution1D-omp, 931100) (dense-embedding-omp, 284100) 
    (entropy-omp, 116) (geodesic-omp, 61010) (heat-omp, 1637750) 
    (jacobi-omp, 4181300) (keogh-omp, 6206200) (layout-omp, 225860) 
    (matrix-rotate-omp, 7950) (michalewicz-omp, 47190) (mixbench-omp, 5953380) 
    (pool-omp, 54140) 
    (randomAccess-omp, 3902310) (stddev-omp, 47560) (stencil1d-omp, 2139240) 
    (vanGenuchten-omp, 457550) (winograd-omp, 14770)
};

\addplot[fill=aclrose, draw=black!80, bar shift=0pt] coordinates {
    (ace-omp, 566420) (atomicCost-omp, 20601070) (bsearch-omp, 12) 
    (colorwheel-omp, 5.7) (convolution1D-omp, 1062940) (dense-embedding-omp, 133340) 
    (entropy-omp, 83) (geodesic-omp, 61230) (heat-omp, 1444190) 
    (jacobi-omp, 8540) (keogh-omp, 148810) (layout-omp, 222110) 
    (matrix-rotate-omp, 1640) (michalewicz-omp, 29690) (mixbench-omp, 1023410) 
    (pool-omp, 46470) 
    (randomAccess-omp, 1204170) (stddev-omp, 4160) (stencil1d-omp, 409330) 
    (vanGenuchten-omp, 299070) (winograd-omp, 5930)
};

\addplot[fill=aclgold, draw=black!80, bar shift=4pt] coordinates {
    (ace-omp, 707030) (atomicCost-omp, 10613640) (bsearch-omp, 25) 
    (colorwheel-omp, 4.3) (convolution1D-omp, 984110) (dense-embedding-omp, 100710) 
    (entropy-omp, 121) (geodesic-omp, 64070) (heat-omp, 1590220) 
    (jacobi-omp, 12820) (layout-omp, 310140) (matrix-rotate-omp, 2590) 
    (michalewicz-omp, 27140) (pool-omp, 77910) 
    (randomAccess-omp, 1037380) (stddev-omp, 26760) (stencil1d-omp, 189560) 
    (vanGenuchten-omp, 576760) (winograd-omp, 170)
};

\legend{Reference, ParaCodex, baseline}

\end{axis}
\end{tikzpicture}
\vspace{-0.5cm}
\caption{\textbf{GPU-time performance of \textsc{ParaCodex} and baseline Codex on HeCBench.}
GPU time (log scale; lower is better) relative to the HeCBench references. Each kernel reports baseline Codex and \textsc{ParaCodex} performance; missing baseline bars indicate failures to produce a correct/compilable implementation. Both systems generally outperform the references, while \textsc{ParaCodex} achieves higher geometric-mean and median gains across the suite.
}

\label{fig:hecbench-gpu-times-compact}
\end{figure*}
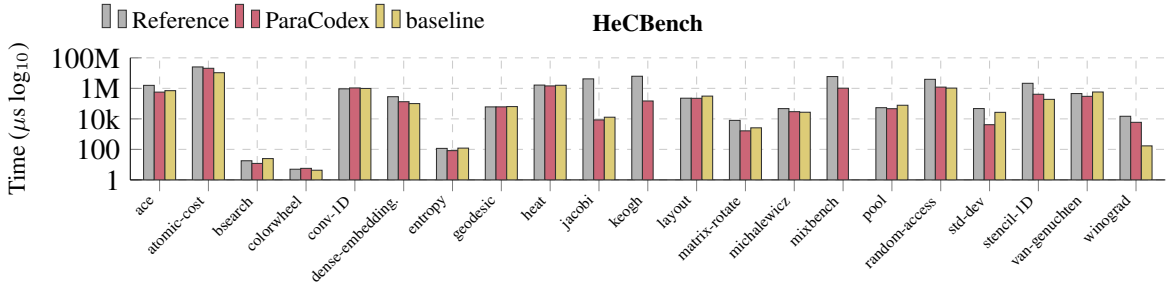
\definecolor{aclrose}{RGB}{204, 102, 119}
\definecolor{aclgold}{RGB}{221, 204, 119}

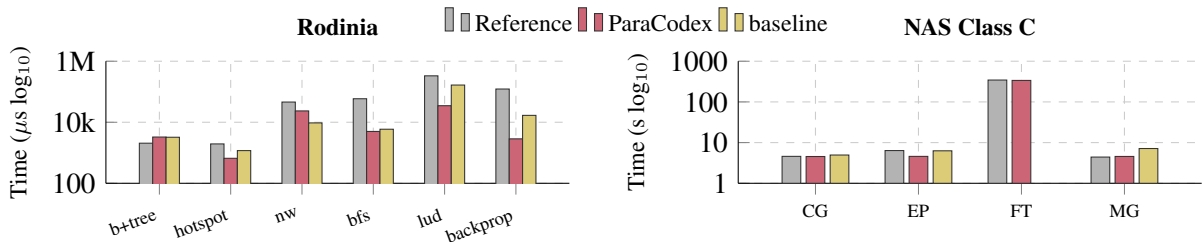
\begin{figure*}[htb]
\centering
\begin{tikzpicture}

\begin{axis}[
    at={(0,0)},
    anchor=south west,
    ybar=0pt,
    ymode=log,
    log basis y=10,
    width=0.48\textwidth,
    height=3.2cm,
    bar width=5pt,
    enlarge x limits=0.15,
    symbolic x coords={b+tree-omp,hotspot-omp,nw-omp,bfs-omp,lud-omp,backprop-omp},
    xtick=data,
    xticklabels={b+tree, hotspot, nw, bfs, lud, backprop},
    x tick label style={
        font=\scriptsize,
        rotate=20,
        anchor=north east
    },
    ymin=100, ymax=1000000,             
    ytick={100,10000,1000000},          
    yticklabels={100,10k,1M},           
    ylabel={Time ($\mu$s log$_{10}$)},
    ylabel style={font=\footnotesize},
    grid=both,
    major grid style={dashed,gray!40},
    minor grid style={dotted,gray!20},
    axis x line*=bottom,
    axis y line*=left,
    title={\textbf{Rodinia}},
    title style={font=\footnotesize, align=center},
    legend style={
        at={(1.15, 1.12)},
        anchor=south,
        legend columns=3,
        font=\footnotesize,
        draw=none
    }
]

\addplot[fill=black!30, draw=black!80, bar shift=-5pt] coordinates {
    (b+tree-omp, 2060) (hotspot-omp, 1950.11) (nw-omp, 45935.959)
    (bfs-omp, 58962.096) (lud-omp, 332150) (backprop-omp, 122780)
};
\addplot[fill=aclrose, draw=black!80, bar shift=0pt] coordinates {
    (b+tree-omp, 3280) (hotspot-omp, 659.036) (nw-omp, 23518.816)
    (bfs-omp, 5000) (lud-omp, 34850) (backprop-omp, 2860)
};
\addplot[fill=aclgold, draw=black!80, bar shift=5pt] coordinates {
    (b+tree-omp, 3226.288) (hotspot-omp, 1177.465) (nw-omp, 9523)
    (bfs-omp, 5907.550) (lud-omp, 165673) (backprop-omp, 16817.154)
};
\legend{Reference, ParaCodex, baseline}
\end{axis}
\begin{axis}[
    at={(0.52\textwidth,0)}, 
    anchor=south west,
    ybar,
    ymode=log,
    log basis y=10,
    width=0.48\textwidth,
    height=3.2cm,          
    bar width=7pt,
    enlarge x limits=0.25,
    symbolic x coords={cg-omp,ep-omp,ft-omp,mg-omp},
    xtick=data,
    xticklabels={CG, EP, FT, MG},
    x tick label style={font=\scriptsize},
    ymin=1, ymax=1000,
    ytick={1,10,100,1000},
    yticklabels={1,10,100,1000},
    ylabel={Time (s log$_{10}$)},
    ylabel style={font=\footnotesize},
    grid=both,
    major grid style={dashed,gray!40},
    minor grid style={dotted,gray!20},
    axis x line*=bottom,
    axis y line*=left,
    title={\textbf{NAS Class C}},
    title style={font=\footnotesize, align=center},
]

\addplot[fill=black!30, draw=black!80] coordinates {
    (cg-omp, 4.591) (ep-omp, 6.375) (ft-omp, 344.812) (mg-omp, 4.410)
};
\addplot[fill=aclrose, draw=black!80] coordinates {
    (cg-omp, 4.553) (ep-omp, 4.586) (ft-omp, 337.606) (mg-omp, 4.569)
};
\addplot[fill=aclgold, draw=black!80] coordinates {
    (cg-omp, 4.951) (ep-omp, 6.285) (mg-omp, 7.153)
};

\end{axis}

\end{tikzpicture}
\vspace{-0.5cm}
\caption{\textbf{Performance of \textsc{ParaCodex} and baseline Codex on Rodinia and NAS reference implementations.}
GPU time over the Rodinia (ms, log scale) \emph{(Left)} and NAS (s, log scale) \emph{(Right)} reference codes. Rodinia includes practical OpenMP programs that are not always fully optimized, while NAS contains highly tuned scientific kernels that serve as a stronger reference point. In both suites, the baseline Codex model already delivers meaningful results. \textsc{ParaCodex} further improves geometric-mean performance on each benchmark. The NAS results demonstrate near-parity with expert implementations.\vspace{-0.5cm}}

\label{fig:combined-gpu-times-compact}
\end{figure*}


\paragraph{Benchmark Suite Evaluation.}

We evaluate \textsc{ParaCodex} across four diverse benchmark suites: ParEval (CUDA$\rightarrow$OpenMP translation fidelity (App.~\ref{app:cuda_migration} details the Serial$\rightarrow$OpenMP to CUDA$\rightarrow$OpenMP migration), \autoref{fig:pareval}), HeCBench (23 diverse micro-kernels, \autoref{fig:hecbench-gpu-times-compact}), Rodinia (application-level benchmarks, \autoref{fig:combined-gpu-times-compact}-left), and NAS (expert-optimized codes, \autoref{fig:combined-gpu-times-compact}-right). \textsc{ParaCodex} demonstrates strong translation reliability: ParEval shows perfect code-only validation (4/4 kernels at 100\%) with strong overall-regime performance (2/4 at 1.0, 1/4 at 0.8); HeCBench achieves 100\% compilation success (21/21 valid kernels after excluding 2 bypass cases ). Performance gains are substantial: HeCBench shows 3$\times$ geometric-mean GPU-time speedup (median 1.59$\times$, 18/21 improved), while Rodinia achieves 5.1$\times$ geometric-mean speedup (median 6.24$\times$). Further comparison with related literature is presented in App.~\ref{app:unipar-comparison}, App.~\ref{app:LASSI-comparison}.

On NAS, \textsc{ParaCodex} matches expert-optimized reference implementations (1.01$\times$ median, 1.08$\times$ geometric-mean speedup). The baseline exhibits lower robustness, failing to compile 2 HeCBench benchmarks and 1 NAS kernel, achieving only 2.4$\times$ (HeCBench) and 3$\times$ (Rodinia) geometric-mean speedups. Detailed per-suite results are provided in App.~\ref{app:detailed-results}, and a code comparison for NAS MG demonstrating 1.57$\times$ improvement over the baseline through profiling-driven kernel fusion are provided in App.~\ref{app:mg-code-example}.

\subsection{Robustness to Anonymization}

\definecolor{aclblue}{RGB}{102,153,204}
\definecolor{aclrose}{RGB}{204,102,119}
\definecolor{aclgold}{RGB}{221,204,119}
\usetikzlibrary{patterns}

\begin{figure*}[!htb]
\centering
\begin{tikzpicture}

\begin{axis}[
    at={(0,0)},
    anchor=south west,
    ybar,
    ymode=log,
    log basis y=10,
    width=0.48\textwidth,
    height=3.2cm,
    bar width=7pt,
    enlarge x limits=0.35,
    symbolic x coords={nw-omp,lud-omp},
    xtick=data,
    xticklabels={nw, lud},
    ymin=100, ymax=1000000,
    ytick={100,10000,1000000},
    yticklabels={100,10k,1M},
    ylabel={Time ($\mu$s log$_{10}$)},
    ylabel style={font=\footnotesize},
    grid=both,
    major grid style={dashed,gray!40},
    minor grid style={dotted,gray!20},
    axis x line*=bottom,
    axis y line*=left,
    tick label style={font=\scriptsize},
    title={\textbf{Anonymized Rodinia}},
    title style={font=\footnotesize, align=center},
    legend style={
        at={(0.45, 1.38)},
        anchor=south,
        legend columns=3,
        font=\footnotesize,
        draw=none
    }
]

\addplot[fill=black!30, draw=black!80] coordinates {
    (nw-omp, 45935.959) (lud-omp, 473283)
};
\addplot[fill=aclrose, draw=black!80] coordinates {
    (nw-omp, 10040.057) (lud-omp, 28376)
};
\addplot[fill=aclgold, draw=black!80] coordinates {
    (nw-omp, 9489) (lud-omp, 46767)
};
\legend{Reference, ParaCodex, baseline}
\end{axis}

\begin{axis}[
    at={(0.52\textwidth,0)},
    anchor=south west,
    ybar,
    bar width=7pt,
    width=0.48\textwidth,
    height=3.2cm,
    ymin=0,
    ymax=700,
    ylabel={Time (s)},
    ylabel style={font=\footnotesize},
    title={\textbf{Anonymized NAS Class D}},
    title style={font=\footnotesize, align=center},
    symbolic x coords={CG,EP},
    xtick=data,
    x tick label style={font=\scriptsize},
    grid=both,
    major grid style={dashed,gray!40},
    minor grid style={dotted,gray!20},
    axis x line*=bottom,
    axis y line*=left,
    enlarge x limits=0.35,
    tick label style={font=\scriptsize},
    label style={font=\scriptsize},
    legend style={
        at={(0.5, 1.38)},
        anchor=south,
        legend columns=4,
        font=\footnotesize,
        draw=none
    }
]


\addplot+[draw=black!70, fill=black!35]
  coordinates {(CG, 607.22) (EP, 95.98)};
\addplot+[draw=black!70, fill=aclrose]
  coordinates {(CG, 563.20) (EP, 89.75)};
\addplot+[draw=black!70, fill=aclblue]
  coordinates {(CG, 429.36) (EP, 90.10)};

\addplot+[draw=black!70, fill=aclgold]
  coordinates {(CG, 0) (EP, 0)};

\legend{Ref., Initial Trans., Targeted Opt., Base.}
\end{axis}

\end{tikzpicture}
\vspace{-0.2cm}
\caption{\textbf{Performance breakdown and generalization.} \emph{(Left)} Anonymized Rodinia GPU execution time (log scale) comparing Reference, \textsc{ParaCodex}, and Baseline on representative kernels. \emph{(Right)} Anonymized NAS Class D performance showing the impact of \textsc{ParaCodex}'s staged refinement (Initial Translation vs.\ Targeted Optimizations) compared to Reference.
}
\label{fig:anonymized-nas-rodinia}
\end{figure*}
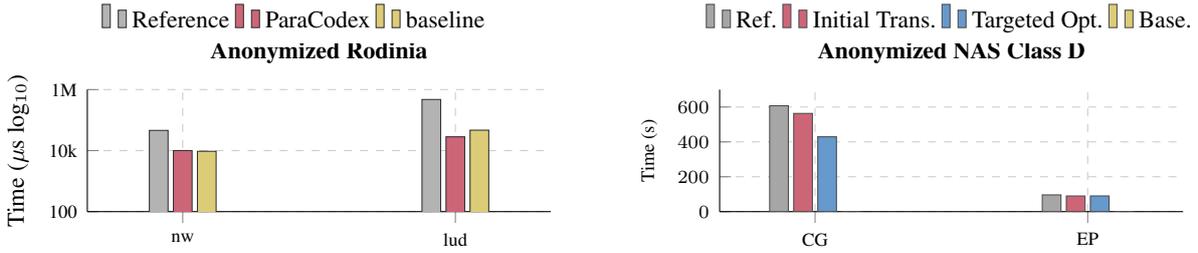
To assess whether \textsc{ParaCodex} depends on memorized identifier patterns, we evaluate anonymized variants of Rodinia and NAS, where all variable and function names are replaced with synthetic tokens and function order is randomized. \autoref{fig:anonymized-nas-rodinia} shows that performance is largely unchanged. On anonymized Rodinia, \textsc{ParaCodex} outperforms the baseline in one varient, and achieves comparable results on the other. On NAS, we evaluated the staged workflow, which also remains effective: Initial Translation establishes strong baselines, and Targeted Optimizations extract further gains (e.g., CG improves to 1.41$\times$). These results indicate that performance stems from structural reasoning and profiler feedback rather than lexical memorization.

\subsection{Behavioral Analysis: Structured Reasoning Under Agentic Control}
\input{latex/thimking_steps_graph}

We further examine how \textsc{ParaCodex} shapes the model’s reasoning during parallelization. \autoref{fig:merged_timeline} compares baseline and \textsc{ParaCodex} timelines for the NAS EP kernel. Under \textsc{ParaCodex}, traces more often decompose the task into parallelization planning and performance-oriented refinement (e.g., parallel region selection, data movement, synchronization), whereas the baseline tends to emit code directly. The bottom-right panel shows similar behavior: for EP, most gains arise in Initial Translation through correct offload and data strategy, while CG improves more gradually across both stages.

\section{Discussion}
We revisit our research questions:

\noindent\textbf{RQ1 (Feasibility):} Contemporary agents demonstrate strong capability for translating serial code to OpenMP GPU offload. The baseline  produces correct and reasonably performant implementations. However, robustness degrades on more complex benchmarks: the baseline fails on NAS FT, exhibits lower compilation success rates overall, and produces less optimized code compared to \textsc{ParaCodex} in all suites. While one-pass generation is often sufficient for simpler kernels, complex scientific codebases with intricate data movement, multi-stage algorithms, and tight synchronization constraints benefit substantially from structured, iterative refinement.

\noindent\textbf{RQ2 (Specialization):} Expert-seeded workflows improve results. \textsc{ParaCodex} consistently boosts compilation rates and speedups by decomposing tasks into analysis, planning, and refinement. As evidenced by reasoning traces, this structure forces the model to externalize its plan before coding, shifting focus from surface-level syntax to performance diagnosis. This allows it to catch logic errors and optimize data transfers that the baseline misses.

\noindent\textbf{RQ3 (Performance):} Crafted agents can match experts. \textsc{ParaCodex} achieves substantial geometric-mean speedups on HeCBench and Rodinia, outperforming the majority of reference implementations. On the highly optimized NAS suite, it matches expert performance, demonstrating that profiling-guided refinement can recover substantial headroom. The GPU-offload bypass behavior observed on 2 HeCBench kernels represents a failure mode where performance-driven optimization violates parallelization intent.

\noindent\textbf{RQ4 (Generalization):} The methodology generalizes. On ParEval (CUDA$\rightarrow$OpenMP), \textsc{ParaCodex} maintains high validity with minimal prompt adjustment. This indicates that the core agentic pattern -- analysis, gated translation, and feedback-driven repair -- is agnostic to the source language, applicable to broader parallel translation tasks beyond the primary serial$\rightarrow$OpenMP setting.

\section{Limitations} Evaluation is limited to a single consumer-grade GPU (RTX 4060) without locked clocks, introducing thermal variance that may affect measurement precision. Correctness validation relies on a \emph{correctness-gate agent} that instruments code with gate macros (checksums and norms at key program points) which may still miss subtle numerical stability issues or non-deterministic race conditions that do not manifest in output differences. The GPU-offload bypass cases (2/36 kernels) represent a failure mode where the agent generates CPU-fallback code that satisfies surface requirements but violates parallelization intent; future work should enforce device-execution constraints as hard requirements. Statistical significance of speedups, particularly for modest gains (e.g., NAS 1.08$\times$), should be validated with larger sample sizes in future studies.

\section*{Acknowledgments}
This research was supported by the Pazy Foundation. Computational support was provided by Code Metal.

\bibliography{latex/new_bib}

\appendix
\section{Pipeline Stage Restructuring for NAS and Rodinia}
\label{app:stage-restructuring}

\subsection{Rationale}
When moving from HeCBench micro-kernels to larger NAS and Rodinia benchmarks, we redesigned the optimization workflow to better reflect full-application performance drivers. Micro-kernels typically expose a small number of localized hotspots, making a sequence of narrowly-scoped optimization stages effective. In contrast, NAS and Rodinia kernels frequently require holistic reasoning over loop structure, data lifetime, launch overhead, and correctness constraints at scale.
Early prototypes decomposed optimization into multiple narrowly scoped stages (e.g., separate concurrency and memory/locality passes). The current pipeline consolidates these into two optimization steps that are applied across suites: (i) GPU offload with an explicit \texttt{data\_plan.md} and baseline capture, and (ii) profiling-driven performance tuning with an explicit \texttt{optimization\_plan.md}. In both cases, each stage is augmented with (i) profiler output from the prior stage, (ii) a short summary of previously applied actions, and (iii) system details (GPU/CPU/memory), enabling profiling-driven iterative refinement under a fixed hardware/software stack.

\subsection{HeCBench Workflow}
For HeCBench, \textsc{ParaCodex} uses a four-stage pipeline developed during our initial experimentation phase. The workflow consists of: (i) analysis, (ii) initial GPU offload with basic correctness validation, (iii) concurrency tuning (e.g., collapse directives, thread mapping), and (iv) memory/locality optimization (e.g., data movement, transfer reduction). Unlike the consolidated three-stage workflow later developed for NAS and Rodinia, this approach used shorter, less comprehensive prompts for each stage and decomposed optimization into narrowly-scoped passes. While this resulted in higher token costs due to the additional stage and context accumulation, it reflected our earlier development methodology and was retained for HeCBench to maintain consistency with completed experiments. The micro-kernel nature of HeCBench benchmarks -- often dominated by a single hot loop -- made the staged decomposition effective despite the increased overhead.

\subsection{NAS/Rodinia Workflow: Consolidated Three-Stage Pipeline}
For NAS and Rodinia, \textsc{ParaCodex} uses a refined three-stage pipeline that consolidates optimization into broader, more comprehensive stages. This redesign was motivated by the characteristics of full-application benchmarks, which require holistic reasoning over loop structure, data lifetime, and launch overhead rather than narrowly-scoped passes. The three stages are: (i)~analysis with loop taxonomy, (ii)~GPU offload with explicit data-planning artifact (\texttt{data\_plan.md}), and (iii)~profiling-driven tuning with optimization planning artifact (\texttt{optimization\_plan.md}). Unlike the four-stage HeCBench pipeline, this workflow uses longer, more detailed prompts that integrate concurrency and memory optimizations into a single profiling-driven refinement stage, reducing token costs while improving reasoning depth. Each prompt is parameterized with profiler output, system metadata, and summaries of prior actions, enabling measurement-driven iteration under a fixed hardware/software stack.

\paragraph{Stage 1 (Analysis Phase): Loop Classification and Data Characterization.}
The analysis phase enumerates and prioritizes loops across the benchmark source files, emphasizing code executed in the main timed region. Each loop is classified using a loop taxonomy that encodes parallelization constraints (e.g., dense, sparse/CSR, multi-stage FFT/butterfly, multigrid, histogram/indirect writes, recurrence, reductions, and stencil patterns). The phase also records loop nesting, dependency structure (reductions, stage dependencies, loop-carried recurrences), and data properties (array roles, access patterns, globals, scratch usage), and flags hazards such as atomics, variable bounds, or insufficient iteration counts. This phase produces a structured \texttt{analysis.md} artifact and copies source files into the kernel working directory without semantic modifications.

\paragraph{Stage 2 (GPU Offload + Data Strategy): Correctness with an Explicit Data Plan.}
Instead of a minimal correctness-first offload step, Stage~1 couples correctness with explicit data-lifetime reasoning. The workflow (i) captures a baseline output for subsequent verification, (ii) selects a data-management strategy using a rule-based decision process (e.g., \texttt{target data} regions, asynchronous overlap, or global device-state allocation via \texttt{omp\_target\_alloc} and \texttt{is\_device\_ptr}), and (iii) requires the pipeline to author a \texttt{data\_plan.md} artifact \emph{before} modifying code. The data plan enumerates arrays used in the timed region, identifies functions that must execute on the device, specifies host-to-device and device-to-host transfer timing and expected volumes, and includes strategy-specific correctness checks to prevent common mapping errors. The implementation then follows this plan to offload the required loops/functions while preserving benchmark semantics and passing the golden-serial correctness check.

\paragraph{Stage 3 (Performance Tuning): Profile-Guided Plan with Early Exit.}
Stage~2 consolidates concurrency tuning, memory/locality tuning, and launch-overhead reduction into a single profiling-driven refinement stage. It first re-verifies correctness against the recorded baseline output. It then reads profiling summaries (kernel time, API time, and transfer time) and records key metrics (dominant kernels, launch counts, and transfer volumes). Before applying changes, the pipeline writes an \texttt{optimization\_plan.md} artifact that prioritizes bottlenecks and proposes specific actions (e.g., hoisting data regions, eliminating redundant transfers, inlining helper functions to reduce launch overhead, loop fusion where safe, increasing collapse depth, introducing reductions for sparse inner loops only when beneficial, enforcing serial stage loops for multi-stage algorithms, and micro-optimizations such as \texttt{const}/\texttt{restrict} and caching locals). The stage supports early termination: if the current runtime is within a small tolerance of the estimated optimum inferred from profiling, the pipeline records metrics and restricts changes to low-risk micro-optimizations.

\subsection{Summary of Differences}
\textsc{ParaCodex} employs two distinct workflows tailored to benchmark characteristics and development timeline. The \emph{four-stage HeCBench pipeline} reflects our initial experimental design with narrowly-scoped, shorter prompts decomposing optimization into separate concurrency and memory passes, resulting in higher token costs. The \emph{three-stage NAS/Rodinia pipeline} represents an improved, consolidated approach that integrates optimization into comprehensive stages with explicit artifact-driven planning (\texttt{data\_plan.md}, \texttt{optimization\_plan.md}). This refinement reduces token overhead while substantially improving reasoning depth and optimization effectiveness for full-application benchmarks. Both workflows share the core philosophy of correctness gating and profiler-driven refinement, and critically, \emph{the correctness gate mechanism itself remained unchanged across both approaches} --- all implementations are validated against golden serial output using the same Makefile-based harness and supervisor-driven repair loop. The three-stage pipeline is strictly superior in terms of cost-effectiveness and reasoning quality; the four-stage approach was retained for HeCBench solely to maintain consistency with completed experiments. Across suites, bottleneck profiles differ: HeCBench micro-kernels tend to be kernel-bound with localized hotspots, whereas NAS/Rodinia benchmarks frequently expose transfer overhead, launch costs, and complex data lifetime patterns that benefit from the holistic data-management planning enabled by the improved workflow.

\section{Analysis, Data Transfer, and Optimization Mechanics}
\label{app:analysis-data-opt}

This appendix details how \textsc{ParaCodex} operationalizes (i) loop-level analysis, (ii) data-movement planning and enforcement, and (iii) profiling-driven optimization during serial-to-OpenMP target-offload translation. The pipeline is instantiated via three prompt templates: an \emph{analysis} prompt used prior to translation, an \emph{offload + data-strategy} prompt that implements a correct GPU version under an explicit data plan, and a \emph{performance tuning} prompt that refines the implementation using profiling feedback. Throughout, prompts are parameterized using runtime template variables (e.g., \texttt{\{kernel\_dir\}}, \texttt{\{file\_listing\}}, and build/run commands) filled by the pipeline scripts.

\subsection{Loop Analysis and Offload Target Identification}
\label{app:analysis-data-opt:analysis}

The analysis phase produces a structured \texttt{analysis.md} artifact while keeping the original source files unmodified. It is designed to (1) identify loops in the main timed region, (2) characterize dependence structure to determine offload feasibility, and (3) surface hazards (reductions, atomics, stage dependencies) that must be handled explicitly during translation.

\paragraph{Loop discovery and prioritization.}
The prompt enumerates loop constructs using lightweight source inspection (e.g., searching for \texttt{for}, \texttt{while}, and main compute-loop patterns). Loops are prioritized by their position in the call path: loops executed every iteration of the main compute loop are treated as \emph{CRITICAL/IMPORTANT}, while one-time setup loops are treated as \emph{SECONDARY/AVOID}.

\paragraph{Priority classification by estimated work.}
For each loop, the prompt applies a coarse work model (iterations $\times$ operations per iteration) and assigns one of four priority levels:
\emph{CRITICAL} (dominant or per-iteration $O(N)$ work), \emph{IMPORTANT}, \emph{SECONDARY}, or \emph{AVOID} (setup/IO or trivially small loops). The primary objective is to focus subsequent offload effort on the loops that plausibly dominate runtime.

\paragraph{Loop-type taxonomy and dependence characterization.}
Each loop is assigned a type from a small taxonomy that encodes parallelization constraints:

\begin{itemize}[nosep,topsep=0pt,leftmargin=*,labelsep=.2em]
  \item \textbf{Type A (Dense):} constant bounds, data-parallel structure.
  \item \textbf{Type B (Sparse/CSR):} inner bound depends on outer index; typically parallelize outer loop, with optional inner parallelism.
  \item \textbf{Type C1 (Multi-stage/Iterative):} stage-dependent computations (e.g., butterfly-like patterns); outer loops parallel, stage loop serial.
  \item \textbf{Type C2 (Multigrid/Hierarchical):} level-wise traversal with dependent stages; outer parallelism with stage ordering preserved.
  \item \textbf{Type D (Histogram/Indirect writes):} parallelizable with atomic updates (or structured privatization + merge).
  \item \textbf{Type E (Recurrence):} loop-carried dependencies or block-level synchronization patterns (e.g., \texttt{\_\_syncthreads()} in CUDA); not offload-parallelizable without algorithmic restructuring.
  \item \textbf{Type F (Reduction):} scalar accumulation; parallelizable via reductions.
  \item \textbf{Type G (Stencil):} neighbor access; parallelizable with careful indexing.
\end{itemize}

A special case is handled explicitly: when an \emph{outer} loop iterates over independent samples while an \emph{inner} loop contains sequential RNG/state updates, the analysis marks the outer loop as parallelizable with per-thread RNG replication and marks the inner RNG loop as sequential within each thread.

\paragraph{Data analysis and hazard flags.}
In addition to loop structure, the analysis records data properties needed for correct mapping:
array shapes (flat vs.\ pointer-based), allocation style (static vs.\ dynamic), accessed struct members, and global variables used in the timed region. It flags hazards that must be handled during translation, including variable bounds, required reductions, atomic updates, stage dependencies, RNG usage in timed regions, and small trip counts.

\subsection{Data Movement and Transfer Strategy}
\label{app:analysis-data-opt:data}

Correct and efficient target offload depends strongly on data-lifetime decisions. The first optimization step therefore requires the pipeline to select a data strategy and to produce a \texttt{data\_plan.md} \emph{before} implementing pragmas. This plan serves as an executable specification of which data reside on the device, when transfers occur, and which functions must execute on the device to avoid implicit host-device thrashing.

\paragraph{Strategy selection rules.}
The prompt selects one of three strategies using the loop taxonomy and program structure:

\begin{enumerate}[nosep,topsep=0pt,leftmargin=*,labelsep=.2em]
  \item \textbf{Strategy A (Scoped \texttt{target data} regions):} use \texttt{target data} with explicit \texttt{map(to|from|tofrom|alloc)} clauses; default for most dense/stencil/reduction kernels and multigrid-like cases.
  \item \textbf{Strategy B (Asynchronous overlap):} use \texttt{nowait} and \texttt{depend} to overlap independent transfers and kernels when the computation permits pipelining.
  \item \textbf{Strategy C (Global device state):} allocate persistent device arrays with \texttt{omp\_target\_alloc} and pass them via \texttt{is\_device\_ptr} to eliminate repeated mapping in iterative solvers and multi-stage kernels.
\end{enumerate}

\paragraph{The \texttt{data\_plan.md} artifact.}
The data plan enumerates all arrays used in the timed region and classifies them as \emph{working}, \emph{scratch}, \emph{const}, or \emph{index}. It also lists functions executed in the timed region and assigns each to host or device execution. The plan explicitly specifies (i) one-time device allocations, (ii) host-to-device transfers (timing and volume), (iii) device-to-host transfers (timing and volume), and (iv) whether any transfers occur inside the iteration loop. The plan includes strategy-specific checklists intended to prevent common errors, such as:
missing device versions of helper functions (triggering implicit copies), scratch buffers remaining on the host when using persistent device pointers, or accidental reintroduction of \texttt{map} clauses into a Strategy~C hot path.

\paragraph{Transfer-volume sanity checks.}
The plan records an \emph{expected transfer volume} for the whole execution and marks large deviations as a red flag. Concretely, if measured transfers exceed the plan by more than a small constant factor (e.g., $>2\times$), the pipeline treats this as evidence of incorrect data scoping (e.g., missing offloaded helpers causing repeated transfers) and prioritizes correcting data lifetime before applying kernel-level tuning.

\subsection{Correctness Gating and Baseline Equivalence}
\label{app:analysis-data-opt:correctness}

Each kernel translation is anchored to a reference baseline captured before modification. The first optimization step requires recording baseline output and verifying the translated output against it (e.g., via textual diff on the benchmark's verification markers). Only configurations that satisfy the golden-serial correctness check are retained for performance reporting. This gating is enforced before profiling-driven tuning, ensuring that subsequent optimizations operate on a semantically valid candidate.

When validation fails, the system invokes a supervisor agent that: (i) instruments code with \texttt{gate.h} macros to trace execution state, (ii) diagnoses discrepancies hierarchically (checking host-device memory consistency then kernel logic), and (iii) applies minimal repairs while preserving GPU offload. The supervisor iterates until \texttt{make check-correctness} passes. The prompt forbids CPU-only fallback and performance changes during this phase --- its sole objective is numerical correctness.

\subsection{Profiling-Driven Optimization and Action Library}
\label{app:analysis-data-opt:opt}

The second optimization step performs profiling-driven refinement while holding the selected data strategy fixed. The prompt explicitly forbids changing the data strategy at this stage to avoid confounding performance gains from data-lifetime restructuring with kernel-level optimizations.

\paragraph{Early exit criterion.}
If the current runtime is within a small tolerance (5\%) of an expected optimum inferred from profiling, or if a proposed change regresses performance beyond 10\%, (e.g., kernel-time dominance and low transfer overhead), the pipeline documents the current metrics and restricts itself to low-risk micro-optimizations (such as \texttt{const}, \texttt{restrict}, and caching locals). This reduces unnecessary refactoring when the implementation is already near a profiler-implied ceiling.

\paragraph{The \texttt{optimization\_plan.md} artifact.}
Before applying changes, the pipeline produces an \texttt{optimization\_plan.md} report containing:
runtime, dominant kernel(s), GPU-time breakdown, transfer fraction and volume, and kernel launch counts. The plan also identifies candidate loop fusions (producer--consumer or adjacent loops with identical bounds) and characterizes iteration structure (e.g., iterative solver loops with repeated SpMV/update patterns).
\paragraph{Prioritized bottleneck checklist.}
The tuning prompt encodes a priority-ordered checklist of bottlenecks and corresponding remedies.  Table~\ref{tab:paracodex-artifacts} summarizes the pipeline artifacts and enforcement points:
\begin{itemize}[nosep,topsep=0pt,leftmargin=*,labelsep=.2em]
  \item \textbf{Data-management issues:} hoist data regions, move scratch to device allocations, ensure all timed-region helpers run on device.
  \item \textbf{Launch overhead:} inline helper functions called inside iteration loops; fuse adjacent loops with identical bounds where safe.
  \item \textbf{Hot-kernel inefficiency:} adjust parallel decomposition (e.g., \texttt{collapse}), add \texttt{simd} to innermost loops, and cache index/array values to reduce redundant loads.
  \item \textbf{Sparse inner-loop decision:} keep CSR inner loops serial when average nonzeros per row is small; introduce inner-loop parallelism with reduction only when nonzeros are sufficiently large to amortize overhead.
  \item \textbf{Stage-dependent algorithms (Type C):} enforce serial stage loops; parallelize only outer dimensions to avoid barrier overhead and correctness failures.
  \item \textbf{Over-parallelization:} remove inner parallelism when outer-loop parallelism already saturates the GPU, as indicated by profiling and problem geometry.
\end{itemize}
\begin{table}[t]
\centering
\small
\begin{tabular}{p{0.16\linewidth} p{0.78\linewidth}}
\toprule
\textbf{Phase} & \textbf{Key Outputs and Enforcement} \\
\midrule
Analysis &
\texttt{analysis.md}: loop taxonomy, priorities, dependencies, data hazards; source copied unmodified. \\
Offload + Data Plan &
\texttt{data\_plan.md}: array inventory, function placement, explicit transfers, expected transfer volume; implementation must follow plan; baseline captured and correctness verified. \\
Tuning &
\texttt{optimization\_plan.md}: profiler-driven diagnosis, prioritized actions, early-exit rule; data strategy fixed; final summary with applied/reverted changes. \\
\bottomrule
\end{tabular}
\caption{Artifacts and enforcement points used by \textsc{ParaCodex} to structure analysis, data movement, and profiling-driven optimization.}
\label{tab:paracodex-artifacts}
\end{table}

\paragraph{Summary and provenance of changes.}
Finally, the prompt requires a structured end-of-step summary that records baseline and final metrics, enumerates applied optimizations (including reverted changes that regress performance), and states the primary insight and remaining bottlenecks. This reporting discipline supports reproducibility and provides provenance for performance improvements reported in the evaluation.

\subsection{Running Example: NAS CG Conjugate Gradient Solver Workflow}
\label{app:running-example}

We illustrate the three-stage workflow on the \textit{CG} (Conjugate Gradient) kernel from NAS Parallel Benchmarks Class C, a sparse linear solver using iterative conjugate gradient with sparse matrix-vector multiplication (SpMV) in CSR format.

\paragraph{(i) Hotspot Summary (\texttt{analysis.md}).}
The analysis phase identifies the main benchmark loop (15 iterations, each invoking \texttt{conj\_grad} with 25 internal \texttt{cgit} iterations) and classifies nested loops:
\begin{itemize}[nosep,topsep=0pt,leftmargin=*,labelsep=.2em]
    \item \textbf{Type E (Sequential):} outer benchmark iteration (\texttt{it = 1..NITER}) and inner \texttt{cgit} loop (25 iterations) with loop-carried dependencies on \texttt{rho}/\texttt{beta} -- must execute serially.
    \item \textbf{Type B (Sparse SpMV):} two SpMV kernels computing \texttt{q = A*p} and \texttt{r = A*z} with irregular CSR indexing via \texttt{rowstr}/\texttt{colidx} -- data-parallel across rows, \emph{CRITICAL priority}.
    \item \textbf{Type F (Reductions):} dot products (\texttt{norm\_temp1}/\texttt{norm\_temp2}, \texttt{rho}, \texttt{d}) and final residual norm -- global reductions, \emph{CRITICAL priority}.
    \item \textbf{Type A (Dense SAXPY):} vector updates (\texttt{z}/\texttt{r}/\texttt{p} axpy operations) -- embarrassingly parallel, memory-bound.
\end{itemize}
\textbf{Data:} CSR matrix (\texttt{a[NZ]}, \texttt{colidx[NZ]}, \texttt{rowstr[NA+1]}, $\sim$461MB total) plus five working vectors (\texttt{x}, \texttt{z}, \texttt{p}, \texttt{q}, \texttt{r}, each \texttt{NA+2} doubles). \textbf{Hazards:} sequential \texttt{cgit} iterations prevent outer parallelism; irregular gather pattern in SpMV limits inner parallelism.

\paragraph{(ii) Data Plan (\texttt{data\_plan.md}).}
Strategy A (persistent \texttt{target data}): establish device residency before benchmark loop with \texttt{\#pragma omp target enter data map(to: a[0:NZ], colidx[0:NZ], rowstr[0:NA+1])} and \texttt{map(alloc: x[0:NA+2], z[0:NA+2], p[0:NA+2], q[0:NA+2], r[0:NA+2])}. Working vectors are initialized on-device to avoid host-device transfers. Expected transfers: 461MB H$\rightarrow$D (CSR data) at entry, negligible D$\rightarrow$H (scalar reduction results only), zero array transfers during 15$\times$25 iteration loops. Correctness check: passes validation with \texttt{VERIFICATION SUCCESSFUL} after initial offload.

\paragraph{(iii) Optimization + Gate Result (\texttt{optimization\_plan.md}).}
Profiling shows runtime dominated by 9,883 kernel launches (400 SpMV passes plus separate reduction/update kernels) with 94\% GPU time in \texttt{conj\_grad} and 91.7\% API overhead from launch/synchronization costs. Bottleneck: repeated small kernels for norm reductions and residual computation inflate launch overhead; each operation spawns separate device kernels. Optimization: (i) fuse dual norm reductions (\texttt{norm\_temp1}/\texttt{norm\_temp2}) into single kernel, (ii) combine final SpMV and residual norm loop to reuse \texttt{rowstr}/\texttt{colidx} loads and eliminate one kernel per \texttt{conj\_grad} call, (iii) cache intermediate scalars in registers to avoid redundant global memory access. Result: kernel launches reduced by $\sim$25\%, runtime improved to 2.04s (estimated $\sim$20\% speedup over baseline). Correctness gate: output validates against serial reference; optimization retained.

\subsection{Baseline Prompt Structure}
\label{app:baseline-prompt}
For comparison, the baseline system uses a single unstructured prompt that combines all objectives into one request. The baseline prompt template is:

\begin{quote}
\textbf{Your Task:}
\begin{enumerate}[nosep,topsep=0pt,leftmargin=*,labelsep=.2em]
    \item Translate the code to an OpenMP GPU-offloaded version.
    \item Apply GPU offloading pragmas as needed.
    \item Optimize the code for performance while preserving functionality.
    \item Ensure the code compiles and runs.
    \item Deliver the modified code to \texttt{\{kernel\_dir\}}.
\end{enumerate}

Deliverable: The complete, modified source code in \texttt{\{kernel\_dir\}}

\end{quote}

This single-pass approach combines analysis, translation, and optimization into one undifferentiated request, contrasting with \textsc{ParaCodex}'s staged workflow that emits intermediate artifacts (\texttt{analysis.md}, \texttt{data\_plan.md}, \texttt{optimization\_plan.md}) and enforces correctness gates between stages. The baseline agent has access to the same tools (compiler, profiler, test harness) as \textsc{ParaCodex} but receives no explicit guidance on when or how to use them. In practice, the baseline agent may compile and test the code, but it does not systematically analyze loops, select data strategies, or iteratively optimize based on profiling feedback, as these steps are not prompted.

\section{From Serial$\rightarrow$OpenMP to CUDA$\rightarrow$OpenMP Translation}
\label{app:cuda_migration}
Our system originally targeted CPU programs and introduced GPU acceleration by
automatically translating serial code to OpenMP target offload. This required
analysis of loop nests, classification of parallelization patterns, and the
construction of an explicit device–residency and data–movement plan.

We later extended this framework to support migration from CUDA code to
OpenMP target offload. Although the overall workflow remained structurally
similar, the design goals and constraints changed substantially. This appendix
summarizes the key differences.

\subsection{\quad Change in Problem Setting}

In the Serial$\rightarrow$OpenMP case, the input code is CPU–bound and
parallelism must be \emph{introduced}. In contrast, CUDA code is already
GPU–parallel. The migration task therefore becomes:

\begin{enumerate}[nosep,topsep=0pt,leftmargin=*,labelsep=.2em]
    \item preserving the program's existing GPU execution semantics,
    \item translating CUDA runtime constructs into OpenMP equivalents, and
    \item avoiding accidental reintroduction of host–device transfers or loss
    of data residency.
\end{enumerate}

Thus the analysis stage was extended to identify not only computational loops,
but also CUDA kernels, launch sites, and device–side execution structure.

\subsection{\quad Extended Analysis for CUDA Kernels}

The original analysis focused on CPU loops and their computational cost. For
CUDA input, the analysis phase additionally:

\begin{itemize}[nosep,topsep=0pt,leftmargin=*,labelsep=.2em]
    \item detects \texttt{\_\_global\_\_} kernels and host launch sites,
    \item reconstructs grid and block structure,
    \item identifies grid–stride loop patterns,
    \item analyzes \texttt{atomicAdd}, shared memory usage, and
          \texttt{\_\_syncthreads()},
    \item estimates total device work as
          \emph{grid~$\times$~block~$\times$~iterations}.
\end{itemize}

This allows the system to reason about CUDA execution semantics that must be
preserved under OpenMP.

\subsection{\quad Data Residency and Memory Model Mapping}

In Serial$\rightarrow$OpenMP translation, device–residence is introduced
incrementally and only when beneficial. However, CUDA programs already assume
GPU–resident data and explicit control over allocation and transfer
(\texttt{cudaMalloc}, \texttt{cudaMemcpy}, \texttt{cudaFree}).

The migration pipeline therefore first reconstructs the CUDA memory model and
maps it to OpenMP constructs.

A central design goal is to avoid unintentionally moving data back to the host
during migration.

\subsection{\quad Execution Model and Synchronization Differences}

CUDA exposes block–local shared memory and synchronization via
\texttt{\_\_shared\_\_} and \texttt{\_\_syncthreads()}. OpenMP target offload
does not provide an exact analogue. The migration workflow therefore performs:

\begin{itemize}[nosep,topsep=0pt,leftmargin=*,labelsep=.2em]
    \item conversion of shared memory buffers to thread–private or local arrays,
    \item replacement of \texttt{atomicAdd} with OpenMP atomics or reductions,
    \item splitting or restructuring kernels that rely on
          \texttt{\_\_syncthreads()}.
\end{itemize}

This represents one of the major conceptual extensions relative to the serial
workflow.

\subsection{\quad Kernel Body and Index Mapping}

CUDA kernels are converted into device functions invoked under
\texttt{target teams loop}. Thread indexing logic involving
\texttt{threadIdx}, \texttt{blockIdx}, and \texttt{gridDim} is replaced by
loop induction variables. Grid–stride loops become conventional bounded loops.

All CUDA API calls and CUDA–specific syntax are eliminated to ensure
pure OpenMP execution.

\subsection{\quad Correctness and Baseline Outputs}

Both workflows enforce diff–based correctness. For CUDA migration, the system
first records the original CUDA program output. The OpenMP–offloaded version
must match this baseline, preventing partial–migration or semantic drift.

\subsection{\quad Performance Considerations}

The CUDA$\rightarrow$OpenMP tuning stage introduces migration–specific
bottlenecks, including:

\begin{itemize}[nosep,topsep=0pt,leftmargin=*,labelsep=.2em]
    \item unintended data transfers after migration,
    \item excessive kernel launch counts after barrier removal,
    \item loss of original CUDA grid dimensionality, and
    \item reduced locality after removing shared memory.
\end{itemize}

Optimizations therefore focus on restoring CUDA–like execution patterns where
possible (e.g., persistent device allocations and loop fusion), while remaining
within the OpenMP execution model.

\subsection{\quad Summary of Conceptual Shift}

In summary, the Serial$\rightarrow$OpenMP pipeline teaches the system how to
\emph{introduce} GPU execution. The CUDA$\rightarrow$OpenMP pipeline teaches it
how to \emph{preserve} an existing GPU execution model in a different runtime
environment. This required extending the analysis phase, formalizing CUDA–to–OMP
semantics, and preventing regression in device residency or synchronization
behavior.

\section{Reproducibility and Experimental Details}
\label{app:reproducibility}

\subsection{Technical Details and Token Counts}
\label{app:token-count}
We use \texttt{gpt-5.1-codex-mini} as the primary model for all kernels. For the NAS FT kernel, we escalated to \texttt{gpt-5.1-codex-max} for both \textsc{ParaCodex} and baseline to maintain fairness. Token usage (NAS benchmarks, representative of current pipeline): \textsc{ParaCodex} averages 837,701 tokens per kernel (cg: 776,036; ep: 967,774; ft: 1,031,208; mg: 575,786), while the baseline averages 755,417 tokens (cg: 286,118; ep: 1,379,682; ft: 880,104; mg: 475,765). The 1.11$\times$ ratio reflects \textsc{ParaCodex}'s multi-stage workflow (analysis → offload+data plan → optimization) compared to the baseline's single-pass translation.

\textbf{Why token usage is high:} Both \textsc{ParaCodex} and the baseline operate Codex CLI in non-interactive automation mode, which is designed for scripting and CI/CD workflows where the agent executes all required actions in a single session without human interaction. This mode consumes significantly more tokens than interactive development because: (i) the full codebase context, build environment, and tool outputs must be provided in each agent turn, (ii) the agent maintains complete execution traces for reproducibility, and (iii) compilation errors, test outputs, and profiler reports are included verbatim to enable deterministic diagnosis. This design prioritizes \emph{reproducibility and determinism} -- ensuring that each parallelization attempt is fully auditable and replayable -- over raw token efficiency. Crucially, the baseline exhibits comparable token consumption (755k tokens/kernel) despite its simpler single-pass structure, confirming that the high token count is primarily an artifact of the non-interactive automation mode rather than \textsc{ParaCodex}'s staged workflow. Full prompt templates are available in our repository.

\subsection{GPU-Offload Bypass: Detailed Analysis}
\label{app:bypass}

In the two HeCBench bypass cases (\textit{pathfinder}, \textit{particlefilter}), \textsc{ParaCodex} produced implementations that compile with OpenMP \texttt{target} directives and pass correctness checks, yet effectively bypass GPU computation by executing the primary computation on the host CPU. The zero-shot Codex baseline exhibits the same behavior on \textit{pathfinder}.

\paragraph{Bypass Definition and Detection.}
We classify a run as a \emph{bypass} when inspection of the generated code together with the corresponding profiler report indicates that the dominant computational loop remains on the host, while the device executes only a minimal or identity kernel that does not account for the kernel's substantive work. Because this behavior contradicts the stated objective of GPU parallelization, including such runs would misrepresent success and can inflate performance comparisons without reflecting genuine offload. We therefore exclude bypass kernels from GPU-time statistics, while reporting their occurrence explicitly.

\paragraph{Bypass Occurrence.}
Bypass was observed in 2/23 HeCBench kernels for \textsc{ParaCodex}. No bypass cases were observed in Rodinia, NAS, or ParEval under our evaluation protocol.

\paragraph{Root Cause Analysis.}
The two bypass cases were analyzed by examining agent transcripts and profiling logs. In both cases, the agent generated code that compiles and passes correctness checks but executes primary computation on the host CPU with minimal/identity GPU kernels. Hypothesis: The agent observed high transfer costs relative to computation and implicitly chose CPU execution as faster, violating the GPU-offload intent. Our inspection suggests bypass arises when the agent judges transfer overheads to dominate and opts to keep computation on the host despite emitting syntactically valid offload constructs.

\paragraph{Mitigation Strategies.}
Looking forward, a harness could mitigate bypass by enforcing evidence of device-side work, for example by requiring non-trivial kernel-launch activity, checking device-memory allocation sizes ($>1$MB), or thresholding the fraction of runtime spent in GPU kernels as measured by profiling (e.g., requiring $>50\%$ of time in device kernels). Preliminary testing with minimum device memory allocation and kernel launch count ($>10$) constraints recovers GPU offload for \textit{pathfinder} but requires algorithmic restructuring for \textit{particlefilter}.

\section{Detailed Benchmark Suite Results}
\label{app:detailed-results}

This section provides comprehensive per-suite analysis of \textsc{ParaCodex}'s performance across ParEval, HeCBench, and Rodinia benchmark suites.

\subsection{ParEval: Robust CUDA$\rightarrow$OpenMP Translation and Correctness}
\label{app:pareval-details}

We evaluate \textsc{ParaCodex} on ParEval, which emphasizes translation fidelity for CUDA$\rightarrow$OpenMP migration (App.~\ref{app:cuda_migration} details the CUDA-specific workflow extensions). Each kernel is assessed under a fixed build environment (\emph{code-only}) and in a setting where the agent must also construct the build flow (\emph{overall}). As shown in \autoref{fig:pareval}, \textsc{ParaCodex} attains perfect compilation and validation in the \emph{code-only} regime (4/4 kernels at 100\%), and remains strong in the more demanding \emph{overall} regime: it compiles and validates \textit{nanoXOR} and \textit{microXOR} in all trials (1.0), achieves 0.8 on \textit{XSBench}, and the remaining errors are concentrated in \textit{microXORH} (0.4). The baseline exhibits consistently lower overall success, with the largest drop on \textit{microXORH}. The heatmap further shows that non-agentic approaches struggle significantly, while top-down agentic methods improve but remain inconsistent. Overall, \textsc{ParaCodex}'s structured workflow -- CUDA kernel analysis, memory model reconstruction, and execution-semantics preservation -- generalizes effectively to cross-API translation.

\subsection{HeCBench: Broad Performance Gains Across Diverse Workloads}
\label{app:hecbench-details}

HeCBench spans regular stencil and dynamic-programming kernels through irregular memory- and graph-oriented workloads. We evaluate 23 benchmarks (\autoref{fig:hecbench-gpu-times-compact}); 2 exhibit GPU-offload bypass and are excluded from performance metrics. We report GPU time for the remaining 21 valid kernels. Parallel-friendly kernels exhibit large gains. For \textit{jacobi} ($489\times$), this stems from reducing redundant transfers via persistent device data regions. Bandwidth-limited kernels improve modestly (e.g., \textit{geodesic}: $1.00\times$, \textit{pool}: $1.17\times$). Across the 21 valid kernels, \textsc{ParaCodex} achieves a 3$\times$ geometric-mean GPU-time speedup (median 1.59$\times$). The baseline demonstrates lower robustness, failing to compile 2 benchmarks, achieving 90\% success (19/21) with a comparable geometric mean of 2.4$\times$ and median of 1.74$\times$. \textsc{ParaCodex} achieves 100\% compilation success (21/21) and improves runtime on 18/21 (86\%).

\subsection{Rodinia: Translation Reliability and Systematic Performance Gains}
\label{app:rodinia-details}

\textsc{ParaCodex} successfully translates all Rodinia kernels and attains a $5.1\times$ geometric-mean GPU-time speedup (median $6.24\times$) (\autoref{fig:combined-gpu-times-compact}-left). Because many Rodinia programs contain correct but non-fully-optimized OpenMP code, \textsc{ParaCodex} frequently improves performance by restructuring parallel regions, reducing synchronization, and lowering data-movement overhead. The baseline also produces running implementations for all kernels but achieves only a $3\times$ geometric-mean speedup (median $3.41\times$).

\subsection{NAS: Handling Complex, Expert-Level Codebases}
\label{app:nas-details}

\autoref{fig:combined-gpu-times-compact} summarizes NAS results (right). \textsc{ParaCodex} achieves GPU-time speedup of $1.01\times$ median and geometric-mean of $1.08\times$ relative to the highly optimized reference implementations. EP and FT show the largest headroom, while MG is sensitive to end-to-end transfer overhead. These results indicate \textsc{ParaCodex} can match expert references on several kernels while remaining competitive on highly tuned codes. The baseline achieves a $0.927\times$ median and $0.834\times$ geometric-mean GPU-time speedup and fails to compile FT\@. App.~\ref{app:mg-code-example} shows a detailed comparison of baseline vs \textsc{ParaCodex} outputs on NAS MG, where \textsc{ParaCodex} achieves 1.57$\times$ better GPU time through kernel fusion driven by profiling feedback.

\section{Comparison with Prior Work}
\label{app:prev_paper_comparison_section}

\subsection{UniPar}
\label{app:unipar-comparison}
\autoref{fig:unipar_comparison} illustrates the substantial gains achieved by a continuously-iterating agent in the serial-to-OpenMP translation task, particularly when guided by an explicit execution plan. The figure reports the fraction of HeCBench OpenMP translations that pass validation across the 13 UniPar~\citep{bitan2025unipar} kernels that also appear in our evaluation set. These are kernels whose core computation resides within a single source file.

\subsection{LASSI}
\label{app:LASSI-comparison}
\citet{dearing2025lassi} proposed a successful pipeline for a related---but distinct---task: translating CUDA to OpenMP\@. Their system achieved an 85\% valid translation rate across ten HeCBench kernels when averaged across several base models. Because their work reports performance relative to the original reference implementation, we evaluate our approach using the same metric on this subset of kernels to enable approximate comparison, despite the differing translation objective.

\autoref{fig:lassi_comparison} reports the \textit{Ratio} metric introduced by \citet{dearing2025lassi}, defined as the reference execution time divided by the execution time of the translated kernel. We present the Ratio for kernels translated by the baseline Codex agent and by \textsc{ParaCodex}, alongside the average Ratio reported by LASSI across its models. Although LASSI evaluates on two A100 GPUs and \textsc{ParaCodex} runs on a single RTX~4060~, reference times in each case are measured on the same hardware as the translated kernels, mitigating most hardware-driven bias.

It is important to emphasize that LASSI evaluates CUDA$\rightarrow$OpenMP translation whereas \textsc{ParaCodex} performs serial$\rightarrow$OpenMP translation; therefore, the comparison is not strictly direct. Nevertheless, the large disparities observed---such as the $489\times$ Ratio achieved by the \textit{jacobi} kernel---together with the consistently higher mean Ratio, suggest that a carefully designed, agentic workflow such as \textsc{ParaCodex} can deliver substantial performance advantages over both out-of-the-box Codex and prior automated pipelines.
\definecolor{aclrose}{RGB}{204, 102, 119}
\definecolor{aclgold}{RGB}{221, 204, 119}

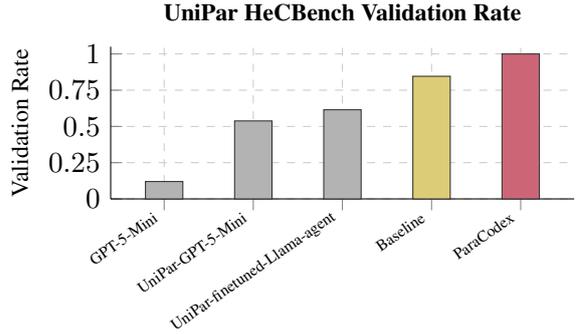
\begin{figure}[t]
\vspace*{0.3cm}
\centering
\begin{tikzpicture}
\begin{axis}[
    ybar,
    bar width=14pt,
    width=0.48\textwidth,
    height=3.6cm,
    ymin=0,
    ymax=1.05,
    ylabel={Validation Rate},
    ylabel style={font=\footnotesize},
    title={\textbf{UniPar HeCBench Validation Rate}},
    title style={font=\footnotesize, align=center},
    symbolic x coords={
        gptmini,
        gptunipar,
        llamaunipar,
        baseline,
        paracodex
    },
    xtick = {
        gptmini,
        gptunipar,
        llamaunipar,
        baseline,
        paracodex
    },
    xticklabels={
        GPT-5-Mini,
        UniPar-GPT-5-Mini,
        UniPar-finetuned-Llama-agent,
        Baseline,
        ParaCodex
    },
    x tick label style={
        font=\tiny,
        rotate=35,
        anchor=east
    },
    ytick={0,0.25,0.5,0.75,1.0},
    grid=both,
    major grid style={dashed,gray!40},
    minor grid style={dotted,gray!20},
    axis x line*=bottom,
    axis y line*=left,
    enlarge x limits=0.15,
]

\addplot[
    draw=black!80,
    fill=black!30,
    bar shift=0pt
] coordinates {
    (gptmini, 0.12)
    (gptunipar, 0.538)
    (llamaunipar, 0.615)
};

\addplot[
    draw=black!80,
    fill=aclgold,
    bar shift=0pt
] coordinates {
    (baseline, 0.846)
};

\addplot[
    draw=black!80,
    fill=aclrose,
    bar shift=0pt
] coordinates {
    (paracodex, 1.0)
};

\end{axis}
\end{tikzpicture}
\caption{\textbf{Percentage of translated code passing validation across different UniPar models}
 Comparing the Codex baseline and \textsc{ParaCodex} against the UniPar methodology with a GPT-5-mini base model and the fine-tuned LLaMA model reported in prior work. The results highlight the substantial validation gains enabled by newer sophisticated agentic models.}
\label{fig:unipar_comparison}
\end{figure}
\definecolor{aclrose}{RGB}{204, 102, 119}
\definecolor{aclgold}{RGB}{221, 204, 119}

\begin{figure}[!t]
\centering
\begin{tikzpicture}
\begin{axis}[
    at={(0,0)},
    anchor=south west,
    ybar=0pt,
    width=0.98\columnwidth, 
    height=3.5cm,
    bar width=4pt,
    enlarge x limits=0.1, 
    symbolic x coords={
        atomicCost-omp, bsearch-omp, colorwheel-omp,
        dense-embedding-omp, entropy-omp, jacobi-omp, 
        layout-omp, matrix-rotate-omp, randomAccess-omp
    },
    xtick=data,
    xticklabels={
        atomicCost, bsearch, colorwheel,
        dense-embedding, entropy, jacobi, 
        layout, matrix-rotate, randomAccess
    },
    x tick label style={
        font=\tiny,         
        rotate=45,
        anchor=north east, 
        inner sep=1pt
    },
    ymin=0, ymax=6,
    ytick={0, 2, 4, 6},
    ylabel={Time rel. to ref.},
    ylabel style={font=\tiny},
    grid=both,
    major grid style={dashed,gray!40},
    minor grid style={dotted,gray!20},
    axis x line*=bottom,
    axis y line*=left,
    title={\textbf{HeCBench}},
    title style={font=\footnotesize, yshift=-1ex},
    legend style={
        at={(0.5, 1.3)}, 
        anchor=south,
        legend columns=3,
        font=\tiny,
        draw=none,
        /tikz/every even column/.append style={column sep=3pt}
    }
]

\addplot[fill=black!30, draw=black!80, bar shift=-4pt] coordinates {
    (atomicCost-omp, 0.5500) (bsearch-omp, 0.1828) 
    (colorwheel-omp, 0.4942) (dense-embedding-omp, 1.0089) 
    (entropy-omp, 0.8051) (jacobi-omp, 3.7455) (layout-omp, 1.0545) 
    (matrix-rotate-omp, 0.3396) (randomAccess-omp, 0.9238)
};

\addplot[fill=aclrose, draw=black!80, bar shift=0pt] coordinates {
    (atomicCost-omp, 1.2284) (bsearch-omp, 1.5) 
    (colorwheel-omp, 0.8771) (dense-embedding-omp, 2.1306) 
    (entropy-omp, 1.3975) (jacobi-omp, 5.9) (layout-omp, 1.0168) 
    (matrix-rotate-omp, 4.8475) (randomAccess-omp, 3.2406)
};

\addplot[fill=aclgold, draw=black!80, bar shift=4pt] coordinates {
    (atomicCost-omp, 2.3845) (bsearch-omp, 0.72) 
    (colorwheel-omp, 1.1627) (dense-embedding-omp, 2.8209) 
    (entropy-omp, 0.9586) (jacobi-omp, 5.9) (layout-omp, 0.7282)
    (matrix-rotate-omp, 3.0694) (randomAccess-omp, 3.7616)
};

\legend{LASSI, ParaCodex, Baseline}

\end{axis}
\end{tikzpicture}
\vspace{-0.3cm}
\caption{\textbf{Reference time / translated kernel run time.} ParaCodex and Baseline for Jacobi are clipped at 6.0 for visualization (actual values: 489.6 -- \textsc{ParaCodex} and 326.1 -- baseline).}
\label{fig:lassi_comparison}
\end{figure}
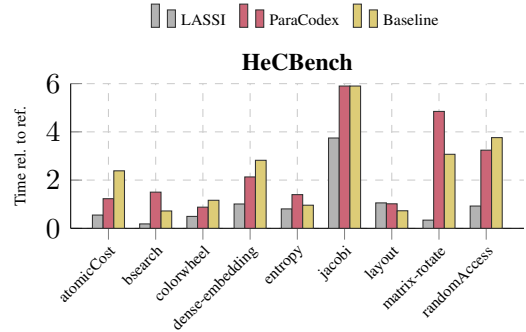

\section{Example Code Transformation: NAS MG vs Baseline}
\label{app:mg-code-example}

To illustrate why \textsc{ParaCodex}\@ outperforms the zero-shot baseline, we compare their outputs on the NAS MG multigrid solver's \texttt{resid} stencil kernel. The baseline produces a correct two-pass implementation with device-allocated temporaries, while \textsc{ParaCodex}'s staged workflow refines this pattern into a fused single-pass structure through profiling feedback.

\subsection{Baseline Output (Zero-Shot Codex)}

The baseline generates a conservative two-pass structure that separates neighbor computations from the residual update:

\textbf{Structure:} The baseline allocates two large temporary arrays (\texttt{u1} and \texttt{u2}, each containing 19.7M doubles, totaling 316MB) on the host using \texttt{malloc}, then maps them to the device using OpenMP's \texttt{map(alloc:)} clause.

\textbf{Pass 1 -- Neighbor sum computation:} A first OpenMP target kernel iterates over the 3D grid (excluding boundary points) and computes 4-neighbor sums along two different axes. For each grid point \texttt{(i3,i2,i1)}, it reads four neighbors from the input array \texttt{ou} and writes the sums to the device-allocated temporary arrays \texttt{u1} and \texttt{u2}. This pass launches 170 times during the solve (once per \texttt{resid} call).

\textbf{Pass 2 -- Residual computation:} A second OpenMP target kernel reads the precomputed sums from global memory temporaries \texttt{u1} and \texttt{u2}, along with the original array \texttt{ou} and the right-hand-side \texttt{ov}, and computes the final residual values into output array \texttt{orr}. This pass also launches 170 times.

\textbf{Performance characteristics:} This implementation incurs (1) host-side \texttt{malloc}/\texttt{free} overhead at each call, (2) 340 total kernel launches (2 per \texttt{resid} call $\times$ 170 calls), (3) redundant global memory traffic (neighbor values written to device memory in pass 1, then read back in pass 2), and (4) implicit synchronization barrier between the two passes. GPU time: 7152 ms (Class C).

\subsection{ParaCodex Output (Staged Workflow)}

\textsc{ParaCodex}'s analysis stage identifies the stencil pattern and flags it as a candidate for fusion. The initial translation stage produces a correct implementation similar to the baseline. The optimization stage then profiles this code, observes high kernel launch counts (340 launches) and detects the memory traffic pattern (write-to-temps followed by read-from-temps), and applies kernel fusion.

\textbf{Structure:} The optimized code eliminates the \texttt{u1} and \texttt{u2} temporary arrays entirely. It uses a single OpenMP target kernel with \texttt{collapse(2)} to parallelize the outer two dimensions.

\textbf{Fused pass -- Register-based computation:} Within the triply-nested loop, the code computes six register variables (\texttt{u1\_c}, \texttt{u1\_L}, \texttt{u1\_R}, \texttt{u2\_c}, \texttt{u2\_L}, \texttt{u2\_R}) that hold neighbor sums computed directly from the input array \texttt{ou}. Each sum aggregates four neighbor values using array accesses with explicit index arithmetic (e.g., \texttt{I3D(i3,i2-1,i1)} for the neighbor below). Immediately after computing these register temporaries, the code uses them to update the residual array \texttt{orr} within the same loop iteration. The register values are never written to global memory.

\textbf{Performance improvements:} This transformation (1) eliminates 316MB device memory allocation and host-side malloc overhead, (2) reduces kernel launches from 340 to 170 (50\% reduction), (3) removes the write-then-read global memory roundtrip for temporary values, and (4) eliminates the implicit barrier between passes. Neighbor values remain in registers throughout, improving cache utilization. \textbf{GPU time: 4569 ms, yielding a 1.57$\times$ speedup over baseline (Class C).}

\subsection{Summary}

This example demonstrates how \textsc{ParaCodex}'s profiling-guided workflow systematically identifies and eliminates performance bottlenecks that single-shot generation misses. The baseline's conservative strategy prioritizes correctness by explicitly materializing intermediate results, but this incurs substantial overhead. \textsc{ParaCodex}'s staged approach allows it to first establish correctness, then use profiling data to safely apply aggressive transformations (kernel fusion, register promotion, launch reduction) that recover 57\% of the baseline's GPU time while maintaining bit-identical output.

\end{document}